\documentclass[a4paper]{qjmam}



\usepackage{graphics,graphicx}
\usepackage{amsmath,amsfonts,amssymb}
\usepackage[dvips]{color}
\usepackage{bm}
\input epsf
\usepackage {mcode}
\def\pmb#1{\setbox0=\hbox{#1}
\kern-.025em\copy0\kern-\wd0
\kern-.05em\copy0\kern-\wd0
\kern-.025em\raise.0433em\box0}

\newcommand{\beq}{\begin{equation}}
\newcommand{\eeq}{\end{equation}}
\newcommand{\be}{\begin{equation}}
\newcommand{\ee}{\end{equation}}
\newcommand{\ba}{\begin{eqnarray}}
\newcommand{\ea}{\end{eqnarray}}

\begin{document}

\title[High-frequency homogenisation for hexagonal and honeycomb lattices]{High-frequency homogenisation for hexagonal and honeycomb lattices}
\author
[M. Makwana and R.~V. Craster]
{M. Makwana and R.~V. Craster}
\address{Department of Mathematics, Imperial College London, South Kensington, London,\\ SW7 2AZ, U.K.}

\received{\recd\today}
\label{firstpage}

\maketitle

\begin{abstract}
A high-frequency asymptotic scheme is generated that captures the motion of waves within discrete hexagonal and honeycomb lattices by creating continuum homogenised equations. The accuracy of these effective medium equations in describing the frequency-dependent anisotropy of the lattice structure is demonstrated. We then extend the general formulation by introducing line defects, often called armchair or zigzag line defects for honeycomb lattices such as graphene, into an otherwise perfect lattice creating surface waves which propagate in the direction of the defect and decay away from it. A quasi-one-dimensional multiple scale method is outlined, which allows us to derive Schroedinger equations describing the local oscillations near particular frequencies in the Bloch spectrum. Further localization by single defects embedded within the line defect are also considered. 

\end{abstract}





\section{Introduction}
\label{sec:intro}

The vibrations of a regular crystal lattice form an essential ingredient of solid state physics and feature heavily in classical texts \cite{brillouin53a,kittel96a,maradudin71a}, with the area of  honeycomb lattices in particular enjoying a renaissance due to modern settings in graphene \cite{neto09a}, honeycomb structures in composites \cite{sparavigna07a}, frame and truss structures \cite{colquitt11a,phani06a}, and in a continuum setting in photonics \cite{zolla05a}. Particularly striking is the dynamic, frequency-dependent, anisotropy of the bulk medium that  leads to exciting and topical applications in optical and acoustic metamaterials \cite{guenneau12a} such as negative refraction, lensing and cloaking. Quite remarkable effects are induced by this effective anisotropy with, at one extreme, all of the energy being concentrated as directional standing waves creating cross shapes of oscillations in both discrete lattice \cite{slepyan08a,osharovich13}, frame \cite{colquitt11a} and photonic \cite{craster12b} systems.

The interpretation and modelling of these dynamic problems is readily performed for perfect lattices using the basic periodic structure to consider an elementary cell that is then repeated to fill space. Much of the behaviour is then interpreted using dispersion curves relating phase shift across the cell to frequency and the resulting iso-frequency contours or Bloch dispersion curves are vital interpretive tools and originate from Brillouin's seminal work \cite{brillouin53a}. Complementary to the study of perfect lattice systems are those containing defects \cite{maradudin65a} or Green's function excitations \cite{barker75a,movchan07a} with exact Green's solutions available for discrete hexagonal, honeycomb \cite{horiguchi72a} or square \cite{economou06a,martin06a} lattice systems. None the less these exact solutions,  given typically as integrals or in elliptic functions, often resist simple interpretation \cite{martin06a} and are complicated by transitions from propagating to stop-band regimes:  It is attractive to alternatively replace a discrete lattice system, or other basically periodic medium, with an effective continuum to avoid the detailed interactions between lattice elements.

For long-wavelength behaviour, that is, when the frequency is low and the wavelength is much greater than the inter-particle spacing, a continuum setting is provided by 
 homogenization theory. This theory takes advantage of the mismatch in scales to create an asymptotic method to upscale from the microscale to the macroscale and is well established \cite{bensoussan78a}. However, most, if not all, of the modern interest and applications are at high frequencies where the wavelength and inter-particle spacings are of similar scale and homogenization theory is no longer applicable. This inadequacy has 
 sparked considerable interest in creating effective continuum models of microstructured media, in various related fields, that break free from the conventional low frequency homogenisation limitations. A suite of extended homogenization theories originating in applied analysis have emerged, for periodic media, called Bloch homogenisation \cite{allaire05a,birman06a,conca95a,hoefer11a}. There is also a flourishing literature on developing homogenized elastic media, with frequency dependent effective parameters, also based upon periodic media as in \cite{nematnasser11a}. Complementary to these is high frequency homogenization \cite{craster10a,guenneau12b} which has had considerable success in modelling effective media for continuous systems in photonics \cite{antonakakis13a} as well as in frames \cite{nolde11a} and elastic plates \cite{antonakakis12a}. All of these applications have been on a square lattice and the homogenization theory was only briefly extended to discrete square lattice systems in \cite{craster10b,makwana13}. Our aim herein is to generalise to the important cases of hexagonal and honeycomb discrete lattice structures thereby creating effective continuum models for them valid away from low frequency.

To further demonstrate the utility of our approach we consider line defects within these regular lattice structures, that is, we consider a hexagonal or honeycomb lattice of identical masses with a single infinite line of altered masses. Surface waves then propagate along the line defect, and decay exponentially perpendicular to the defect, and are analogous to the Rayleigh-Bloch waves that exist for continuous systems \cite{porter99a,wilcox84a}.
For square lattices, with an embedded line defect, such waves are shown to exist \cite{joseph13a} and the dispersion curves found both exactly and through high frequency homogenization; the latter represents the line defect in the continuum setting as an effective string with the lattice and the masses incorporated into effective parameters. Likewise for the hexagonal and honeycomb line defect systems dispersion curves and effective properties are extracted and are relevant to, for instance, the edge states of graphene. 

The plan of this article is as follows: In section 2 we illustrate our homogenisation theory by initially formulating the problem of wave propagation in hexagonal and honeycomb geometries. We begin by considering Bloch waves in perfectly periodic lattices and move on to test the efficacy of our method by comparing our asymptotic solutions with those found using numerical simulations on lattices which have had an external force applied on the scale of the microstructure. In section 3 we introduce line-defects into our otherwise perfect lattice structure which lead to Rayleigh-Bloch waves. For the honeycomb lattice both zigzag and armchair defects are investigated using our method, where for the former we further demonstrate the two-scale approach for the case of a lattice containing a single defective mass within the embedded line defect. Finally in section 4 some concluding remarks are drawn together.

\section{Discrete lattices}
\label{sec:first}
We begin by considering perfect lattice structures treating both hexagonal and honeycomb lattices as these are intimately connected. A crucial detail is that the micro-structure has a very clear and natural representation in a local short-scale lattice coordinate system which is not orthogonal, whilst on the global long-scale an orthogonal Cartesian coordinate system is natural.

\subsection{Hexagonal lattices}
\vspace{0.2cm}
\subsubsection{Formulation}
\begin{figure}[h!]
\begin{center}
    \includegraphics[height=10cm, width=12.5cm]{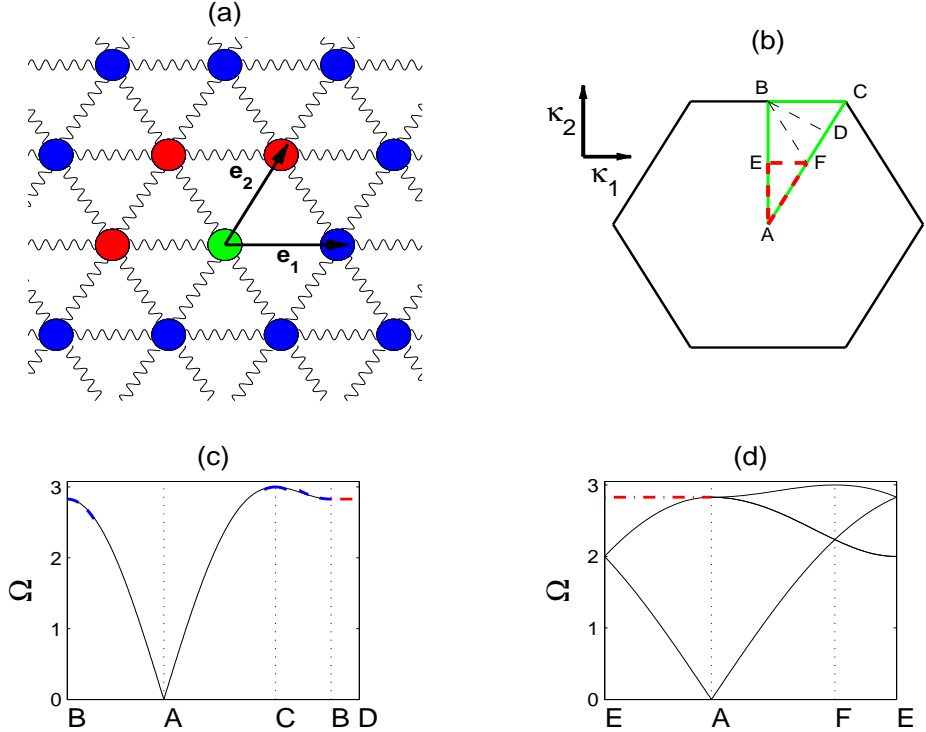}
\end{center}
\caption{\small {Panel (a) displays the hexagonal mass-spring lattice where the lattice basis vectors are indicated by ${\bf e}_1$ and ${\bf e}_2$. Panel (b) shows the hexagonal 1st Brillouin zone, where the irreducible zone for the single mass cell is indicated by the larger triangle $\text{A}=\left(0,0\right)$,     $\text{B}=\left(0,2\pi/\sqrt{3}\right)$,  $\text{C}=\left(2\pi/3,2\pi/\sqrt{3}\right)$, whilst the irreducible zone for the 4 mass elementary cell is shown by the dashed red lines $\text{E}=\left(0,\pi/\sqrt{3}\right)$,  $\text{A}$,  $\text{F}=\left(\pi/3,\pi/\sqrt{3}\right)$. Panels (c) and (d) shows the dispersion diagram for the single mass and 4 mass elementary cell, respectively. The  dashed lines in panel (c) were derived from the asymptotic method, equations \eqref{eq:triangle_pde} and \eqref{eq:triangle_pde2}. The dot-dash flat portion of the dispersion diagram in (c,d) corresponds to wavenumbers along the line $\text{B}$,  $\text{D}=\left(\pi/2,\pi\sqrt{3}/2\right)$ shown in (b). }}
\label{fig:triangle_dispersion}
\end{figure}

We initially consider wave propagation through a uniform hexagonal lattice (figure \ref{fig:triangle_dispersion}(a)) where we consider transverse oscillations to the plane; this is sometimes called a triangular lattice \cite{horiguchi72a}.  Our two-scale approach will generate an effective equation which describes the macroscale motion explicitly whilst implicitly detailing the microstructure oscillations in the coefficients. With this objective in mind, we define the short-scale discrete coordinates $n,m$ along the lattice basis vectors ${\bf e}_1$ and ${\bf e}_2$ respectively, as shown in figure \ref{fig:triangle_dispersion}(a). 

\hspace{-0.35cm} As we are primarily concerned with long-scale wave propagation through the lattice, with wavelength potentially on the scale of the microstructure, we assume that the distance between the masses is $\epsilon \ll 1$; the basis vectors are written as ${\bf e}_1=\epsilon {\bf i}$ and ${\bf e}_2=\epsilon \left( 1/2{\bf i}+\sqrt{3}/2 {\bf j} \right)$ where ${\bf i}$ and ${\bf j}$ are unit vectors in an orthogonal Cartesian coordinate frame. This representation of the lattice vectors provides a connection 
between the microstructure and macrostructure as the long-scale coordinates, which we denote by $\eta_1,\eta_2$ and treat as continuous, are based on the orthogonal coordinate system
\beq
\eta_1=\epsilon \left(n+\frac{m}{2} \right) \hspace{0.5cm} \eta_2=\epsilon \left(\frac{\sqrt{3}m}{2} \right).
\label{eq:triangle_lattice_transformation} \eeq 
Our aim is to obtain effective partial differential equations posed entirely upon this long-scale, but which none the less capture the short-scale features within coefficients. 

Our analysis begins with a simple model expressed by the following non-dimensional difference equation 
\beq
-M \Omega^2 y_{n,m}=y_{n+1,m}+y_{n-1,m}+y_{n,m+1}+y_{n,m-1}+y_{n-1,m+1}+y_{n+1,m-1}-6 y_{n,m},
\label{eq:triangular_difference} 
\eeq
 where $y_{n,m}$ denotes the displacement of the mass situated at ${\bf r}=n {\bf e}_1+m {\bf e}_2$, $\Omega$ is the vibration frequency and $M$ is the mass. Note that the governing equation is equivalent to a second-order accurate $7$ point approximation to the Laplacian on a hexagonal grid for the wave equation \cite{zak2005} considering nearest neighbour interactions only; additionally lengths, tension and other parameters have been scaled out of the problem. Throughout this paper we only consider time-harmonic motion, hence the multiplicative factor $\exp\left(-i\Omega t \right)$ (where $t$ is the time component) is considered understood and suppressed henceforth. For a defect-free hexagonal lattice, the phase shift between masses is represented by the Bloch wavenumber vector $\mbox{\boldmath$\kappa$}=\left( \kappa_1,\kappa_2 \right)$, where $\mbox{\boldmath$\kappa$}= \kappa_1 {\bf i} + \kappa_2 {\bf j}$ and the quasi-periodicity condition is defined as
\beq
y_{n+\hat{N},m+\hat{M}}= \exp \left[i\left(\hat{N} \kappa_1 + \frac{\hat{M}}{2} \left[\kappa_1+\sqrt{3}\kappa_2\right] \right) \right] y_{n,m}.
\label{eq:triangle_bloch} \eeq
 Substitution into the difference equation gives the well-known dispersion relation relating frequency to phase shift as 
\beq
\Omega^2=\frac{1}{M} \left[6-4 \cos\left(\frac{\kappa_1}{2} \right)\cos\left(\frac{\kappa_2 \sqrt{3}}{2} \right) -2 \cos\left(\kappa_1 \right) \right],
\label{eq:triangle_dispersion_relation} \eeq 
the dispersion curves are shown in figure \ref{fig:triangle_dispersion}(c).

In \cite{craster12b} it was shown that, for a square lattice, there exists an interesting mode of oscillation, which is missed if the usual route of  plotting the dispersion relation (\ref{eq:triangle_dispersion_relation}) along the edges of the irreducible Brillouin zone is chosen. It is interesting to note that this missing mode is also present for the hexagonal lattice and is formed by traversing the diagonal path, shown as BD in figure \ref{fig:triangle_dispersion}(b). This mode corresponds to the flat band in the dispersion curve (figure \ref{fig:triangle_dispersion}(c)) and implies zero-group velocity for a specified range of $\mbox{\boldmath$\kappa$}$. The flat-band is notably more prevalent if we consider an elementary cell comprised of 4 masses (figure \ref{fig:triangle_dispersion}(d),  $0 \le \kappa_1 \le \pi/3$ and $0 \le \kappa_2 \le \pi/\sqrt{3}$); visually, this is demonstrated by the way that the smaller triangle is reflected within the larger one of figure \ref{fig:triangle_dispersion}(b). It is anticipated that this missing mode is present within all two-dimensional Bravais lattices in addition to specific non-Bravais structures such as the periodic honeycomb lattice that will be considered later. The oscillations associated with this flat band and the highly anisotropic response found is analysed later using the asymptotic technique (\ref{eq:forced_hyperbolic_equation}) and related to the star-waves found in \cite{slepyan08a}. 

\subsubsection{Asymptotics}
Employing the asymptotic method of \cite{craster10b}  to the hexagonal mass-spring model allows us to derive an equation which encapsulates the motion on the long-scale with the microscale behaviour implicitly defined. This is achieved by treating the long-scale and the short-scale coordinates as independent, hence the displacement 
\beq
y_{n+N,m+M}= y(\eta_1+\widehat{\eta_1},\eta_2+\widehat{\eta_2},N,M),
\label{eq:representation}
\eeq
in this setting where $\widehat{\eta_1}=\epsilon \left(N+M/2 \right)$ and  $\widehat{\eta_2}=\epsilon \left(\sqrt{3}/2 \right)M$: naturally we view our macroscale in an orthogonal coordinate frame, \eqref{eq:triangle_lattice_transformation}, whilst the short-scale oscillations take into explicit consideration the hexagonal lattice geometry.  Hence the first two independent variables in (\ref{eq:representation}) are associated with the continuous macroscale motion along $\mbox{\boldmath$\eta$}=\eta_1 {\bf i}+\eta_2 {\bf j}$; the latter two variables correspond to the discrete microscale motion along the hexagonal lattice basis vectors, ${\bf e}_1$ and ${\bf e}_2$. If one is concerned merely with the perfect lattice and Bloch problem  then the Bloch relation \eqref{eq:triangle_bloch} applied to the displacement function (\ref{eq:representation}) has the natural separation that 

\beq
y(\eta_1+\widehat{\eta_1},\eta_2+\widehat{\eta_2},N,M)= \exp \left[i\left(N \kappa_1 + \frac{M}{2} \left[\kappa_1+\sqrt{3}\kappa_2\right] \right) \right] y(\eta_1+\widehat{\eta_1},\eta_2+\widehat{\eta_2},0,0).
\label{eq:triangle_bloch_asy} \eeq 
This is indicative that the purely continuous displacement function $y(\eta_1+\widehat{\eta_1},\eta_2+\widehat{\eta_2},0,0)$ is key; we omit the last two arguments hereon $y(\eta_1+\widehat{\eta_1},\eta_2+\widehat{\eta_2})$, which is expanded in orders of $\epsilon$ as 
\[
y(\eta_1+\widehat{\eta_1},\eta_2+\widehat{\eta_2})=\]
\beq y(\mbox{\boldmath$\eta$})+\widehat{\eta_1} \frac{\partial y(\mbox{\boldmath$\eta$})}{\partial \eta_1}+ \widehat{\eta_2} \frac{\partial y(\mbox{\boldmath$\eta$})}{\partial \eta_2}+ \frac{1}{2} \left(\widehat{\eta_1} \frac{\partial^2 y(\mbox{\boldmath$\eta$})}{\partial \eta_1^2}+\widehat{\eta_2} \frac{\partial^2 y(\mbox{\boldmath$\eta$})}{\partial \eta_2^2}+2\widehat{\eta_1}\widehat{\eta_2} \frac{\partial^2  y(\mbox{\boldmath$\eta$})}{\partial \eta_1\partial \eta_2}\right) +\mathcal{O}(\epsilon^3).
\label{eq:triangle_taylors}
\eeq 
To apply the asymptotic method we expand both the continuous displacement function and the frequency squared in powers of $\epsilon$
\beq
 y\left(\eta_1,\eta_2 \right)= y_0\left(\eta_1,\eta_2 \right)+\epsilon y_1\left(\eta_1,\eta_2 \right)+\epsilon^2 y_2\left(\eta_1,\eta_2 \right) + O(\epsilon^3), 
\label{eq:y_expansion}
\eeq
\beq \Omega^2=\Omega_0^2+\epsilon\Omega_1^2+\epsilon^2\Omega_2^2+O\left(\epsilon^3 \right).
\label{eq:triangular_displ_expansion} \eeq 
This separation of scales of the displacement, and the subsequent expansions, are applied to the difference equation \eqref{eq:triangular_difference} with the resulting equations solved in orders of $\epsilon$.
\\ 

Our initial interest is in deriving a continuous equation which characterises motion near the flat band frequency at point $B$, $\left(0,2\pi/\sqrt{3} \right)$. Following the methodology of \cite{craster10b,makwana13} the standing wave frequency is $\Omega_0=\sqrt{8/M}$,  $\Omega_1=0$, and the second-order correction leads to the long-scale equation for $y_0$ as
\beq
3\frac{\partial^2 y_0}{\partial \eta_2^2}-\frac{\partial^2 y_0}{\partial \eta_1^2} -2M \Omega_2^2 y_0=0 .
\label{eq:triangle_pde} \eeq 
We check the validity of this for a perfect lattice, using the quasi-periodic Bloch condition about $\mbox{\boldmath$\kappa$}=\left(0,2\pi/\sqrt{3}\right)$,
\beq
y_0=f_0 \exp\left[i \left(\eta_1 k_1+\eta_2\left(k_2-\frac{2\pi}{\sqrt{3}\epsilon} \right)\right)\right].
\eeq  
This is substituted into equation \eqref{eq:triangle_pde} which generates the asymptotic dispersion relation $\Omega_2^2=k_1^2/2M$, where $(\kappa_1,\kappa_2)=\epsilon (k_1,k_2)$ and $f_0$ is an arbitrary constant ($\mbox{\boldmath$\kappa$}:B\rightarrow C$ in figure \ref{fig:triangle_dispersion}(c)). 

The methodology works well around all the standing wave frequencies at the edges of the Brillouin zone, for instance around the point $C$ in the vicinity of the highest point of the dispersion curve, $\Omega_0=\sqrt{9/M}$ the long-scale equation 
\beq
\nabla^2 y_0 -\frac{4}{3} M \Omega_2^2 y_0=0,
\label{eq:triangle_pde2} \eeq
 where $\nabla^2=\partial_ {\eta_1}^2+\partial_{\eta_2}^2, \partial_{\eta_j}=\partial/\partial \eta_j$ for $j=1,2$, is found. Recalling that $\Omega_2^2$ is the perturbation from the standing wave frequency we see that if this correction is negative then this is a Helmholtz equation allowing propagating solutions and conversely if this is positive we expect decaying solutions; this is in agreement with intuition from the dispersion curves. We compare this long-scale equation and the Green's function lattice forcings in section \ref{sec:forcing}.

\subsubsection{Fourier transform and numerics}
A canonical example is the Green's function for the hexagonal lattice and we modify the system by forcing the lattice: the left-hand side of equation \eqref{eq:triangular_difference} acquires an additional $-F \delta_{n,0} \delta_{m,0}$ forcing term. 
We define the semi-discrete Fourier transform, and its inverse, as 
\beq
Y(k_1,k_2)=\sum_{n\in\mathbb{Z}}\sum_{m\in\mathbb{Z}} y_{n,m} \exp\left(i \left[{\bf k}.\mbox{\boldmath$\eta$}(n,m) \right] \right),
\label{eq:forwardDFT}\eeq
\beq
y_{n,m}=\frac{1}{2a b} \int\limits_{-a}^{a} \int\limits_{0}^{b} Y(k_1,k_2) \exp\left(-i \left[{\bf k}.\mbox{\boldmath$\eta$}(n,m) \right] \right) d{\bf k},
\label{eq:backwardDFT}\eeq where ${\bf k}=k_1{\bf i} + k_2{\bf j}, a=2\pi/\left(\sqrt{3}\epsilon\right)$ $b=2\pi/\epsilon$ and  $\mbox{\boldmath$\eta$}(n,m)=\eta_1 {\bf i}+\eta_2{\bf j}$ describes the nodal positions.  Applying this transform to the amended difference equation gives the solution as 
\beq
y_{n,m}=-\frac{F}{2a b \epsilon^2}\int\limits_{-a\epsilon}^{a \epsilon} \int\limits_{0}^{b \epsilon} \frac{\exp\left(-i \left[\kappa_1 \left(m/2+n \right) + \kappa_2 \left(\sqrt{3}m/2 \right) \right] \right)}{M\Omega^2-6+2\cos\left(\kappa_1 \right)+4\cos\left(\kappa_1/2 \right)\cos \left( \sqrt{3}\kappa_2/2 \right)} d \kappa_1 d \kappa_2 .
\label{eq:exact_integral}
\eeq 
As expected, the denominator of the function in the integral \eqref{eq:exact_integral} is the dispersion relation \eqref{eq:triangle_dispersion_relation}. If $\Omega>\sqrt{9/M}$, we obtain a decaying defect mode in the non-propagating region of the Bloch diagram. In that case, 
 known integrals from \cite{gradshteyn07a} reduce (\ref{eq:exact_integral}) to a single integral
\beq
y_{n,m}=-\frac{F}{\pi}\int\limits_{0}^{\pi}\frac{\cos\left[\left(2n+m\right)\zeta \right]}{A\sqrt{1-a^2}}\left(\frac{\sqrt{1-a^2}-1}{a} \right)^{|m|} d\zeta ,
\label{eq:exact_single_integral}\eeq where
\beq
a=B/A, \hspace{0.5cm} A=\Omega^2-6+2\cos\left(2\zeta \right), \hspace{0.5cm} B=4 \cos(\zeta).
\nonumber
\eeq 
In the propagating region the  lattice Green's function solutions have been obtained 
 in terms of elliptic integrals for the hexagonal cases \cite{horiguchi72a}.

It is convenient to have an efficient, and independent, numerical 
 alternative, and check, upon our results. Truncating the infinite system to a finite system, $N\times N$, of masses (the lattice contain $N$ masses in the ${\bf e_1}$ and ${\bf e_2}$ direction) and reformulating the forced variation of the governing equation \eqref{eq:triangular_difference} results in the following matrix equation  
\beq
D{Y}+{Y}D+E{ Y}E+E^T{Y}E^T+M \Omega^2 Y=-F.
\label{eq:matrix_equation_triangle} 
\eeq 
Here $D, E$ are sparse matrices of size $N\times N$ being zero everywhere except along specific diagonals; matrix $D$ consists of $-3$ along the main-diagonal with $1$ in both off-diagonal positions, $E$ merely contains $1$ along a single off-diagonal (the super-diagonal), $F$ is the forcing matrix which contains the forcing value in the central position and $Y$ corresponds to the matrix of displacements. Numerically, we solve the above equation by transforming it into a large $N^2\times N^2$ matrix-vector problem by utilising the Kronecker product. There is the natural question of which boundary conditions to employ and we utilize a variant of perfectly matched layers (PML), as outlined in \cite{makwana13}, to prevent spurious reflections from the edges of the domain.

\subsubsection{Forcing and comparison}
\label{sec:forcing}
Given the exact solution, and the matrix approach, of the previous section we proceed to see how the asymptotic solution fares. We begin near the edge of the Brillouin zone, at the point C, where $\mbox{\boldmath$\kappa$}=\left( 2 \pi/3, 2\pi/\sqrt{3} \right)$; we augment \eqref{eq:triangle_pde2} by incorporating the forcing. The forcing is moved to the long-scale using $\delta_{n,0}\delta_{m,0}=\epsilon^2 \delta(\eta_1)\delta(\eta_2)$ and the asymptotic governing equation is
\beq
\nabla^2 y_0 -\frac{4}{3} M \Omega_2^2 y_0=-\frac{4}{3}F\delta(\eta_1)\delta(\eta_2).
\label{eq:forced_asymptotic_equation_stop_band} 
\eeq 
 Solutions to \eqref{eq:forced_asymptotic_equation_stop_band} are the Bessel function Green's function to the Pseudo-Helmholtz equation which are
\beq
y_0=-\frac{2 F}{3 \pi} K_0\left(2\sqrt{\frac{M}{3}}\Omega_2|{\bf r}| \right),
\label{eq:hankel_solution_t} 
\eeq 
where $|{\bf r}|=\sqrt{\eta_1^2+\eta_2^2}$. In \cite{makwana13} we considered square lattices  and both elliptic and hyperbolic equations of a similar form to this hexagonal case were obtained; the solutions for forced square lattices were also Bessel functions.

\begin{figure}
\begin{center}
    \includegraphics[height=5cm, width=12.5cm]{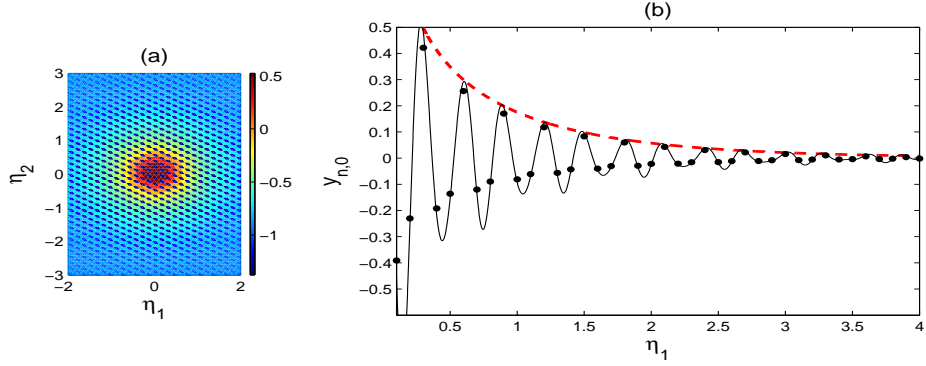}
\end{center}
\caption{\small {In panel (a) the long-scale axisymmetric nature of the lattice displacements, at a frequency close to $\Omega^2=9/M$, close to point C, in figure \ref{fig:triangle_dispersion}(b) is demonstrated. Panel (b) shows a comparison between the asymptotic solution \eqref{eq:hankel_solution_t}, the dashed envelope, the numerics from the matrix equation \eqref{eq:matrix_equation_triangle} (solid points) and the solution from \eqref{eq:exact_single_integral} (solid line). The comparison is taken  along the $\eta_1$ direction ($m=0$). In both panels, the frequency is $\Omega^2=9/M+0.005$ where $M=1$, for the matrix approach the number of points in the lattice is $N=87$ and $\epsilon=0.1$.}}
\label{fig:triangle_comparison2}
\end{figure}
\begin{figure} [htb]
\begin{center}
    \includegraphics[height=6cm, width=11.5cm]{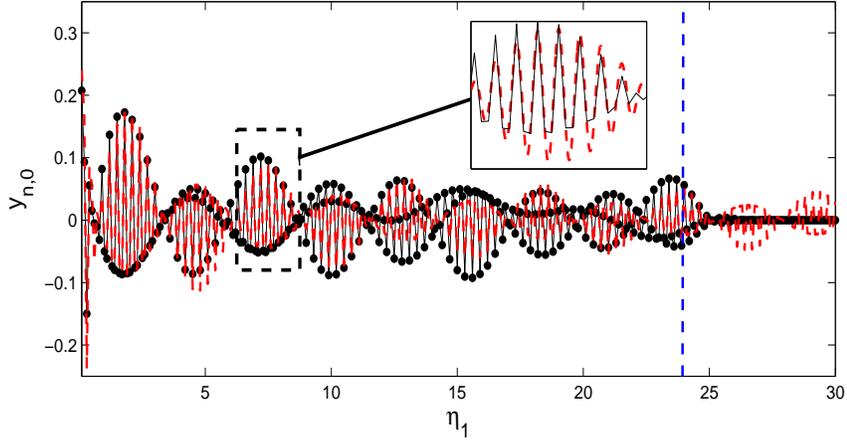}
\end{center}
\caption{\small {A direct comparison between the solution from matrix equation \eqref{eq:matrix_equation_triangle} (shown by the solid line/points) and the asymptotic solution \eqref{eq:hankel_solution_t} (dashed envelope) for a frequency within the propagating region. The dashed oscillating lines are derived by combining  the long-scale modulation with the short-scale oscillations \eqref{eq:triangle_bloch} plotted at the discrete lattice points. The comparison is along the $\eta_2=0$ direction. We have chosen the frequency as $\Omega^2=9/M-\epsilon^2, \Omega_2^2=-1, \epsilon=0.1$, the mass value is $M=1$ and, for the matrix approach, the number of lattice points is $N=601$, where we have applied a layer of PML $61$ (the vertical dashed line indicates the start of PML) masses deep from the edge in $(n,m)$ space. A closer examination of both solutions for a limited range of $\eta_1$ is shown in the inset.}}
\label{fig:triangle_comparison3}
\end{figure}

We now compare the Bessel function solution \eqref{eq:hankel_solution_t} to the $y_{n,m}$ derived using the Fourier transforms \eqref{eq:exact_single_integral} and the numerical solution of the matrix problem \eqref{eq:matrix_equation_triangle}. In the decaying region
typical results are shown in figure \ref{fig:triangle_comparison2}. The comparison between the numerical results and the asymptotic solution about point C within the propagating region is  shown in figure \ref{fig:triangle_comparison3}. In both examples shown, the lattice consists of $601^2$ masses in $(n,m)$ space where we have applied a surrounding layer of PML $61$ masses deep and in both cases the asymptotics perform very well capturing the long-scale decay and oscillations respectively. 

\begin{figure}
\begin{center}
    \includegraphics[height=5cm, width=7.75cm]{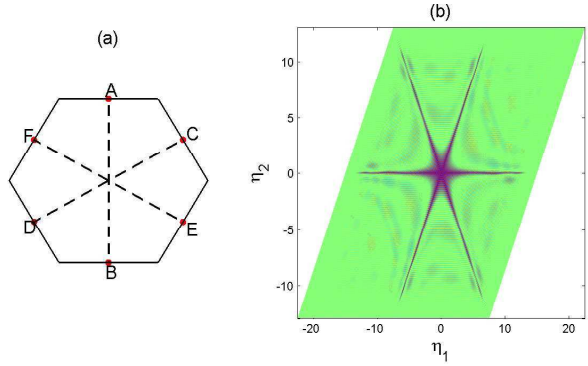}
\end{center}
\caption{\small {Panel (a) shows the first Brillouin zone where the points associated to the hyperbolic PDE's  of the form \eqref{eq:forced_hyperbolic_equation} are labelled. Panel (b) shows the star-like oscillation, derived from the forced variation of the  matrix equation \eqref{eq:matrix_equation_triangle}, for fixed frequency $\Omega^2=\sqrt{8}-0.01$. The lattice has been truncated to contain $N=301$ masses in the ${\bf e}_1,{\bf e}_2$ direction, with PML applied $31$ masses deep around the perimeter, the mass value is taken as $M=1$. }}
\label{fig:x_star4}
\end{figure}

\indent We now move onto analysing the forced variation of the PDE \eqref{eq:triangle_pde} which governs motion near the flat-band at point B, 
\beq
3\frac{\partial^2 y_0}{\partial \eta_2^2}-\frac{\partial^2 y_0}{\partial \eta_1^2} -2M \Omega_2^2 y_0=-F \delta\left(\eta_1 \right)\delta\left(\eta_2 \right).
\label{eq:forced_hyperbolic_equation} 
\eeq 
An important point versus \eqref{eq:forced_asymptotic_equation_stop_band} is that the equation has changed character to become hyperbolic; 
 this equation has solutions 
\beq 
 y_0( \eta_1,\eta_2) =\begin{cases} \mathcal{A}K_0\left(\Omega_2 \gamma \sqrt{2M/3}  \right) & \gamma > 0 \\ \mathcal{B} H_0^{(1)}\left(\Omega_2 \gamma \sqrt{2M/3} \right) & \gamma < 0 \end{cases},
 \label{eq:flat_band_t_sol}\eeq where $\gamma^2=\eta_2^2-3\eta_1^2$ and $\mathcal{A},\mathcal{B}$ are constants. An important feature of the solution is that that along ${\bf e}_2$ and ${\bf e}_2-(1/2){\bf e}_1$ axes we obtain a logarithmic singularity. For the square lattices treated in \cite{makwana13} a hyperbolic equation of the form \eqref{eq:forced_hyperbolic_equation} gave lattice oscillations predominantly along the two characteristics of the associated PDE. For the hexagonal lattice,  if the lattice is excited at the frequency $\Omega\left(0,2\pi/\sqrt{3} \right)$ we obtain star-like oscillations (figure \ref{fig:x_star4}(b)) along three characteristics, not just the X-wave oscillations found in the square case along two characteristics. The three characteristics are justified by examining the PDE's analogous to \eqref{eq:forced_hyperbolic_equation} which describe standing-wave oscillations at the points indicated in figure \ref{fig:x_star4}(a). Points A,B correspond to the PDE \eqref{eq:triangle_pde} whilst the solution of the equations  at points C-D and E-F indicate that the characteristics are $\pm \pi/3$ rotations of those at points A-B. Hence when the lattice is excited at the flat band frequency we obtain a star-like pattern. These line-localised waveforms, for the hexagonal and square mass-spring lattices are also discussed in \cite{slepyan08a}, and for lattice frames \cite{colquitt12a}; these star-like waveforms also occur for continuous media \cite{craster12b} and even for structured elastic media \cite{antonakakis13b} and are a generic feature of waves in microstructured media.

\subsection{Honeycomb lattice}
\begin{figure}
\begin{center}
    \includegraphics[height=6.75cm, width=11.5cm]{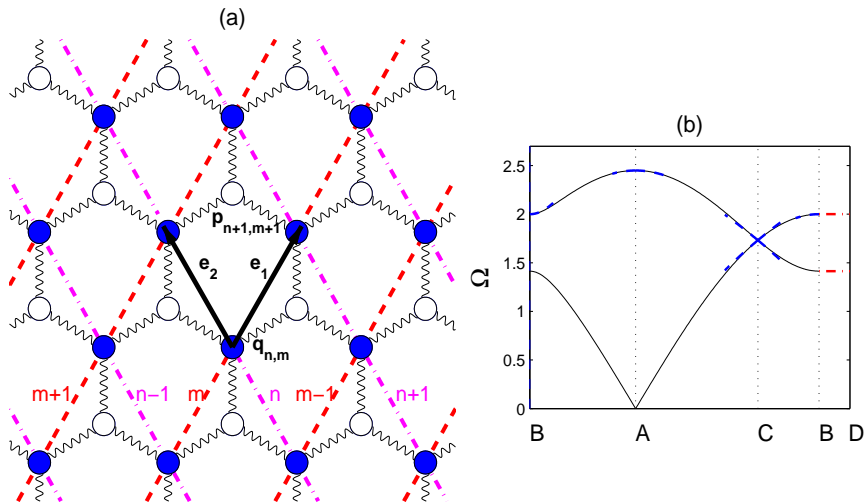}
\end{center}
\caption{\small {Panel (a) shows the honeycomb mass-spring model where $p_{n,m}$ represents the displacement of the white masses whilst $q_{n,m}$ indicates the displacement of the blue masses. The magenta dash-dot lines represent integer values along the ${\bf e}_1$ axis whilst the dashed lines correspond to integer values along the ${\bf e}_2$. Panel(b) shows the dispersion curve for the honeycomb lattice, where $M=1$, and where we have used the irreducible Brillouin zone BAC shown in figure \ref{fig:triangle_dispersion}(b). The asymptotics, almost indistinguishable from the exact solution (solid lines), are indicated by the blue dashed lines whilst the red dash-dot lines shows the hidden mode, which corresponds to the line BD within the irreducible zone (figure \ref{fig:triangle_dispersion}(b)).}}
\label{fig:honeycomb_dispersion}
\end{figure}

Honeycomb lattices, figure \ref{fig:honeycomb_dispersion}, contain the same basis vectors as the hexagonal lattice and hence are closely Mathematically related, there is however an important nuance which is that there are now coupled difference equations to consider.

\subsubsection{Formulation}
 The formulation of the honeycomb lattice, figure \ref{fig:honeycomb_dispersion} (a), follows a similar vein to that of the hexagonal lattice, where the long-scale orthogonal coordinate $\mbox{\boldmath$\eta$}$ is now defined as
\beq 
 \eta_1=\frac{1}{2}\epsilon \left(n-m \right) \hspace{0.5cm} \eta_2=\frac{\sqrt{3}}{2}\epsilon \left(n+m \right), 
\label{eq:honeycomb_lattice_transformation} 
\eeq
 $n,m \in \mathbb{Z}$ are defined along the honeycomb lattice basis vectors ${\bf e}_1=\epsilon\left(1/2 {\bf i}+\sqrt{3}/2{\bf j} \right),{\bf e}_2=\epsilon\left(-1/2 {\bf i}+\sqrt{3}/2{\bf j} \right)$ (as shown in figure \ref{fig:honeycomb_dispersion}(a)) and $\epsilon \ll 1$. The non-dimensional form of the difference equations governing motion in the honeycomb lattice are now a coupled system 
\beq
-\Omega^2 M p_{n,m}=q_{n,m-1}+q_{n-1,m}+q_{n,m}-3p_{n,m}, \hspace{0.2cm}-\Omega^2 M q_{n,m}=p_{n,m}+p_{n,m+1}+p_{n+1,m}-3q_{n,m},
\label{eq:honeycomb_difference_equations1} 
\eeq 
where  $M$ is the mass value at the locations associated to $q_{n,m},  p_{n,m}$, as shown in figure \ref{fig:honeycomb_dispersion}(a). An arbitrary position within the lattice is defined as ${\bf r}=n {\bf e}_1+m {\bf e}_2$  and the lattice basis vectors are shown in figure \ref{fig:honeycomb_dispersion}(a). A similar Bloch periodicity condition to equation \eqref{eq:triangle_bloch} holds for the honeycomb structure
 \beq
{\bf y}_{n+N,m+M}=\exp \left(i \left[\frac{N}{2} \left(\kappa_1+\sqrt{3}\kappa_2 \right)+\frac{M}{2}\left(\sqrt{3}\kappa_2-\kappa_1 \right) \right] \right) {\bf y}_{n,m},
\label{eq:honeycomb_bloch} 
\eeq 
where $N,M \in \mathbb{Z}$ and ${\bf y}_{n,m}$ represents the displacement vector ${\bf y}_{n,m}=\left[p_{n,m}, q_{n,m} \right]^T$. The dispersion relation is derived by substituting the above Bloch relation into the difference equations and solving accordingly, thereby giving
\beq
\Omega_{O,A}^2=\frac{3}{M} \pm\sqrt{\frac{1}{M^2}+\frac{4}{M^2}\cos\left(\frac{\kappa_1}{2}\right)\left[\cos\left(\frac{\kappa_1}{2}\right)+\cos\left(\frac{\sqrt{3}\kappa_2 }{2}\right)\right]},
\label{eq:honeycomb_dispersion} 
\eeq
$\Omega_O$ represents the optical mode ($+$) and $\Omega_A$ the acoustic mode ($-$). 
The honeycomb lattice in figure \ref{fig:honeycomb_dispersion}(a) is a non-Bravais structure that is constructed using a basis, consisting of two masses, replicated over the entire lattice at identical locations to the masses in the hexagonal structure. Our formulation resembles the phenomenological nearest-neighbour tight-binding model of graphene,  where the primitive basis vectors are identical to that of the hexagonal lattice, hence it follows that the Brillouin zone of the honeycomb lattice is given by figure \ref{fig:triangle_dispersion}(b). Due to the similarities in the composition of the hexagonal and honeycomb lattice, the hidden mode appears in an identical location in $\mbox{\boldmath$\kappa$}$ space, as shown in the dispersion diagram of figure \ref{fig:honeycomb_dispersion}(b).

\subsubsection{Asymptotics}
\label{sec:hexagonal_asymptotics}
The coupled difference equations \eqref{eq:honeycomb_difference_equations1} are treated in a similar manner to the earlier hexagonal lattice case; we again define the macroscale in terms of the orthogonal coordinates $\mbox{\boldmath$\eta$}=\eta_1 {\bf i}+\eta_2 {\bf j}$ and the short-scale oscillations using the primitive lattice basis vectors. Similarly to the hexagonal lattice asymptotics of section \ref{sec:hexagonal_asymptotics} we utilise the Bloch relation \eqref{eq:honeycomb_bloch} to show the natural separation of scales 
\beq
p(\eta_1+\widehat{\eta_1},\eta_2+\widehat{\eta_2},N,M)= \exp \left(i \left[\frac{N}{2} \left(\kappa_1+\sqrt{3}\kappa_2 \right)+\frac{M}{2}\left(\sqrt{3}\kappa_2-\kappa_1 \right) \right] \right) p(\eta_1+\widehat{\eta_1},\eta_2+\widehat{\eta_2},0,0),
\label{eq:hexagon_bloch_asy} 
\eeq
c.f. \eqref{eq:triangle_bloch_asy}, 
 where $\widehat{\eta_1}=\epsilon/2(N-M)$ and $\widehat{\eta_2}=\epsilon\sqrt{3}/2(N+M)$; $q(\eta_1+\widehat{\eta_1},\eta_2+\widehat{\eta_2},N,M)$ is treated in an identical manner. The continuous displacement is conveniently written as a vector function, with first component  $p(\eta_1+\widehat{\eta_1},\eta_2+\widehat{\eta_2},0,0)=p(\eta_1+\widehat{\eta_1},\eta_2+\widehat{\eta_2})$ and second component  $q(\eta_1+\widehat{\eta_1},\eta_2+\widehat{\eta_2},0,0)=q(\eta_1+\widehat{\eta_1},\eta_2+\widehat{\eta_2})$. The transformation \eqref{eq:hexagon_bloch_asy} is applied the difference equations and we Taylor expand as in \eqref{eq:triangle_taylors}; the frequency is expanded in powers of $\epsilon$, as in \eqref{eq:triangular_displ_expansion},  and the ansatz 
\beq
p\left(\eta_1,\eta_2 \right)=p_0\left(\eta_1,\eta_2 \right)+\epsilon p_1\left(\eta_1,\eta_2 \right)+\epsilon^2 p_2\left(\eta_1,\eta_2 \right) + O(\epsilon^3), \eeq
\beq
q\left(\eta_1,\eta_2 \right)=q_0\left(\eta_1,\eta_2 \right)+\epsilon q_1\left(\eta_1,\eta_2 \right)+\epsilon^2 q_2\left(\eta_1,\eta_2 \right) + O(\epsilon^3).
\label{honeycomb_displ_expansion} 
\eeq  
 is applied for $p$ and $q$. 

This leads to a series of equations in orders of $\epsilon$, valid at the various standing wave frequencies,  that we solve order-by-order. Following the two-scale methodology of \cite{craster10b,makwana13} we derive the continuous long-scale equation valid in the vicinity of the point $A$ $\bm\kappa=(0,0)$, this PDE governs motion about the highest point in the dispersion curve. The leading order (standing wave) frequency and the equations relating the displacement functions are found to be 
\beq
\Omega_0=\sqrt{6/M}, \hspace{0.5cm} {\bf y}_0= \left[ -1, 1 \right]^T q_0 \left(\eta_1,\eta_2 \right),\quad p_1=-\left(\frac{1}{2}\left[\frac{\partial p_0}{\partial \eta_1}+\frac{1}{\sqrt{3}}\frac{\partial p_0}{\partial \eta_2} \right] +q_1 \right),
\label{eq:highestpoint_honeycomb}\eeq
\beq
\nabla^2q_0\left(\eta_1,\eta_2 \right)-4M \Omega_2^2 q_0\left(\eta_1,\eta_2 \right)=0.
\label{pointA_leading_order_solution_honeycomb}
\eeq 
We can also derive the leading order frequency and displacement function relations, valid near the flat-band frequency at point $B$, $\bm\kappa \left(0,2 \pi/\sqrt{3} \right)$,
\beq
\Omega_0=\sqrt{4/M}, \hspace{0.5cm} {\bf y}_0= \left[ 1, 1 \right]^T p_0 \left(\eta_1,\eta_2 \right), \hspace{0.2cm} p_1=-\left(\frac{3}{2}\left[\frac{\partial p_0}{\partial \eta_1}+\frac{1}{\sqrt{3}}\frac{\partial p_0}{\partial \eta_2} \right] -q_1 \right),
\label{eq:flatband_honeycomb}\eeq
\beq
\frac{\partial^2 p_0}{\partial \eta_1^2}-3\frac{\partial^2 p_0}{\partial \eta_2^2}-4 \Omega_2^2 M p_0\left(\eta_1,\eta_2 \right)=0.\label{eq:pointB_leading_order_solution_honeycomb}
\eeq
The PDE in \eqref{eq:pointB_leading_order_solution_honeycomb} is hyperbolic and     
 if we excite the lattice at the flat-band frequency we obtain star-like oscillations  similarly to that for the hexagonal lattice  in figure \ref{fig:x_star4}(b). This hyperbolic equation indicates that oscillations occur in an X-shape, predominately along the directions of the ${\bf e}_1$ and ${\bf e}_2$ axis, for both sets of masses associated to the displacements $p_{n,m}$ and $q_{n,m}$. When we excite the lattice at the flat-band frequency we obtain oscillations formed from the superposition of these X-waves.

An important feature of the dispersion curves is that there is the locally linear crossing of the dispersion  curves at the point C: the so-called Dirac point. The absence of a finite stop-band is because of these Dirac points,  located at the six corners of the first Brillouin zone; if we had a diatomic honeycomb structure, with alternating masses within a single hexagonal cell then a finite stop-band would open up, \cite{Fefferman12}. When we apply our multiple scale scheme at the Dirac point $\bm\kappa=(2\pi/3,2\pi/\sqrt{3})$, we get the leading order frequency and displacement components  
\beq
\Omega_0=\sqrt{3/M}, \hspace{0.5cm} {\bf y}_0(\eta_1,\eta_2)=\left(-\left[\sqrt{3}/\left(2M\Omega_1^2 \right)\right] \left[\partial_{\eta_2} -i\partial_{\eta_1} \right], 1 \right) q_0\left(\eta_1,\eta_2 \right),
\label{eq:pointC_honeycomb_eqn2}\eeq where $q_0$ is found from solving following elliptic equation
\beq
\nabla^2q_0\left(\eta_1,\eta_2 \right)+\frac{4}{3}M^2 \Omega_1^4 q_0\left(\eta_1,\eta_2 \right)=0,
\label{eq:pointC_honeycomb_eqn}\eeq which is derived using the solvability condition at $\mathcal{O}(\epsilon)$. 

\subsubsection{Fourier transform and numerics}
In order to stringently verify our asymptotic method for the honeycomb lattice, we shall apply a localised external forcing to the central mass associated to the $q_{n,m}$ displacements and compare the resulting solutions. For completeness we initially outline the derivation of a precise solution using Fourier transforms, although due to ease of computation, we shall opt to validate our multiple-scale scheme against the numerics. 

Initially we use the semi-discrete Fourier transform on $p_{n,m},q_{n,m}$, as defined in \eqref{eq:forwardDFT} and \eqref{eq:backwardDFT},where $\mbox{\boldmath$\eta$}(n,m)=(\eta_1,\eta_2)$ with $\eta_j$ defined in \eqref{eq:honeycomb_lattice_transformation} .We apply this transform to the forced variation of the difference equations, whereby there is an additional $-F \delta_{n,0}\delta_{m,0}$ term on the left side of the second difference equation in \eqref{eq:honeycomb_difference_equations1}. After some algebra, the resulting equations are solved to find the coupled displacements 

\beq
q_{n,m}=-\frac{F}{2a b \epsilon^2}\int\limits_{-a \epsilon}^{a \epsilon} \int\limits_{0}^{b \epsilon} \frac{\exp\left(-i/2 \left[\kappa_1 \left(n-m \right) + \sqrt{3}\kappa_2 \left(n+m \right) \right] \right)\left(M\Omega^2-3 \right)}{M^2\Omega^4-6M\Omega^2+6-4\cos\left(\kappa_1/2 \right)\cos\left(\sqrt{3}\kappa_2/2 \right)-2\cos\left(\kappa_1 \right)} d \kappa_1 d \kappa_2 ,
\label{eq:exact_integral_honeycomb_q}
\eeq 
which is reduced to 
\beq
q_{n,m}=-\frac{F(M \Omega^2-3)}{2\pi}\int\limits_{0}^{\pi}\frac{\cos\left[\left(n-m\right)\zeta \right]}{A\sqrt{1-a^2}}\left(\frac{\sqrt{1-a^2}-1}{a} \right)^{|n+m|} d\zeta ,
\label{eq:exact_single_integral2}\eeq where
\beq
a=B/A, \hspace{0.5cm} A=M^2\Omega^4-6M\Omega^2+6-2\cos\left(2\zeta \right), \hspace{0.5cm} B=4 \cos(\zeta),
\nonumber\eeq 
and the displacements $p_{n,m}$ are found by solving the following integral
\beq
p_{n,m}=-\frac{F}{2a b \epsilon^2}\int\limits_{-a\epsilon}^{a \epsilon} \int\limits_{0}^{b \epsilon}\frac{Q(k_1,k_2)\left[2\cos\left(\kappa_1/2\right)\exp\left(i\kappa_2\sqrt{3}/2 \right)+1 \right]}{M\Omega^2-3}  d \kappa_1 d \kappa_2,
\label{eq:exact_integral_honeycomb_p}\eeq where $Q\left(k_1,k_2 \right)$ denotes the Fourier transform of \eqref{eq:exact_integral_honeycomb_q}. 

As in the hexagonal case a useful alternative matrix representation is found by truncating the lattice at some fixed value $N$ along the ${\bf e}_1$ and ${\bf e}_2$ directions, and solving the ensuing matrix equations (similar to equation \eqref{eq:matrix_equation_triangle} for the hexagonal lattice). The difference equations \eqref{eq:honeycomb_difference_equations1}, incorporating forcing at the location associated to the displacement $q_{0,0}$, are formulated as
\beq
D^T Q+Q D+(M\Omega^2-3)P=0, \hspace{0.5cm}D  P+P D^T+(M\Omega^2-3)Q=-F,
\label{eq:honeycomb_matrix_equations} 
\eeq 
where $D$ consists of $1$'s along the superdiagonal and $1/2$ along the main diagonal; $F$ corresponds to the forcing, whereby a forcing term is in the central position of the matrix and $P,Q$ are the displacement matrices. Solving the first equation for $P$, substituting into the second, and simplifying gives
\beq
HQ+QH+EQE+E^TQE^T=(M\Omega^2-3)F, \hspace{0.4cm} H=G+T-\frac{(M\Omega^2-3)^2}{2}I_N,
\label{eq:honeycomb_matrix_equations2}
\eeq
 where $G$ consists of $3/2$ along the main diagonal and $1$ along both off-diagonals, above and below the main diagonal;$T$ has a single entry of $-1$ in the $(N,N)$'th position of the matrix. The first equation of \eqref{eq:honeycomb_matrix_equations2} is solved, as in the hexagonal lattice matrix equation \eqref{eq:matrix_equation_triangle}, with the resulting displacement vector $Q$ substituted back into the first equation of \eqref{eq:honeycomb_matrix_equations}, thereby allowing us to derive the solution for the displacements $p_{n,m}$. Once again the effects of truncating the domain are ameliorated using discrete PMLs as described in \cite{makwana13}.

\subsubsection{Forcing and comparison}
\begin{figure}[h!]
\begin{center}
    \includegraphics[height=5cm, width=11cm]{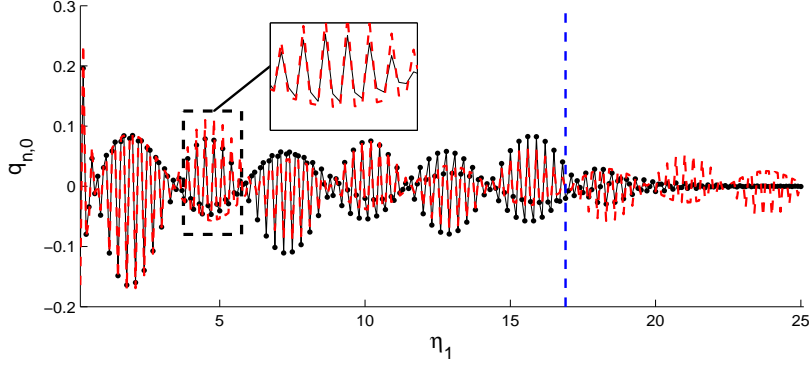}
\end{center}
\caption{\small {This figure shows a comparison of $q_{n,0}$ between the asymptotic and matrix method near the Dirac point in the dispersion curve for the honeycomb lattice. The displacements are plotted for the mass value $M=1$, where the beginning of the active PML range is indicated by the dashed vertical line. The asymptotics are shown by the dashed oscillating lines, whilst the solid curve and points are derived from the matrix method. The inset shows a closer examination of the two solutions for a specified range of $\eta_1$. }}
\label{fig:hexagon_comparison3}
\end{figure}

\begin{figure} [h!]
\begin{center}
    \includegraphics[height=5cm, width=11cm]{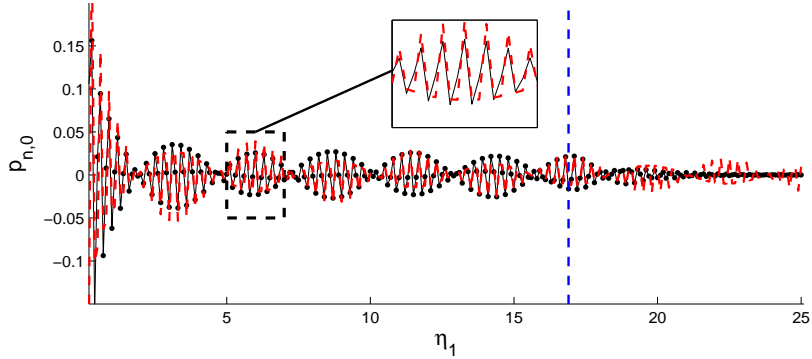}
\end{center}
\caption{\small {This figure shows the comparison for the $p_{n,0}$ displacements.}}
\label{fig:hexagon_comparison4}
\end{figure}

The validity of the multiple-scale method for the honeycomb structure is demonstrated with a  comparison between the matrix  and asymptotic method, for the $p_{n,m}$ and $q_{n,m}$ displacements, in the the pass-band, figure \ref{fig:hexagon_comparison3} and \ref{fig:hexagon_comparison4} (a comparison in the decaying region is trivial due to the corresponding standing wave frequency being located at the origin in reciprocal space), hence we shall opt to focus on the local oscillations about the Dirac point. When we force a single mass located at the position associated to $q_{0,0}$, the second difference equation in \eqref{eq:honeycomb_difference_equations1} is amended with a $-F\delta_{n,0}\delta_{m,0}/\epsilon$ term on the left-hand side. The previous equation \eqref{eq:pointC_honeycomb_eqn} becomes inhomogeneous with an additional $-4M\Omega_1^2F\delta\left(\eta_1 \right)\delta\left(\eta_2 \right)/3$ term on the right-hand side, whereby after solving we obtain the following solution 
\beq
q_0(\eta_1,\eta_2)=-iM\Omega_1^2\left(F/3 \right) H_0^{(1)} \left[2\sqrt{M/3}\Omega_1^2 |{\bf r}| \right].
\eeq Note that the long-scale displacement $p_0(\eta_1,\eta_2)$ satisfies an anisotropic equation \eqref{eq:pointC_honeycomb_eqn2}, hence the solution compared to that of the matrix method (figure \ref{fig:hexagon_comparison4}) is deduced by taking the superposition of the functions at each of the conical points within the first Brillouin zone. 

\section{Surface waves}
In this section we shall demonstrate the existence of Rayleigh-Bloch waves in hexagonal and honeycomb lattices with line defects, whereby waves propagate along the defect and exponentially decay in the opposing direction. Our method will be utilised to show local frequency variation about wavevectors, where the lattice oscillates in a standing wave pattern. Throughout this section we shall prescribe the defect mass value such that it is less than the bulk lattice mass value. 
\begin{figure}[h!]
\begin{center}
    \includegraphics[height=5cm, width=10cm]{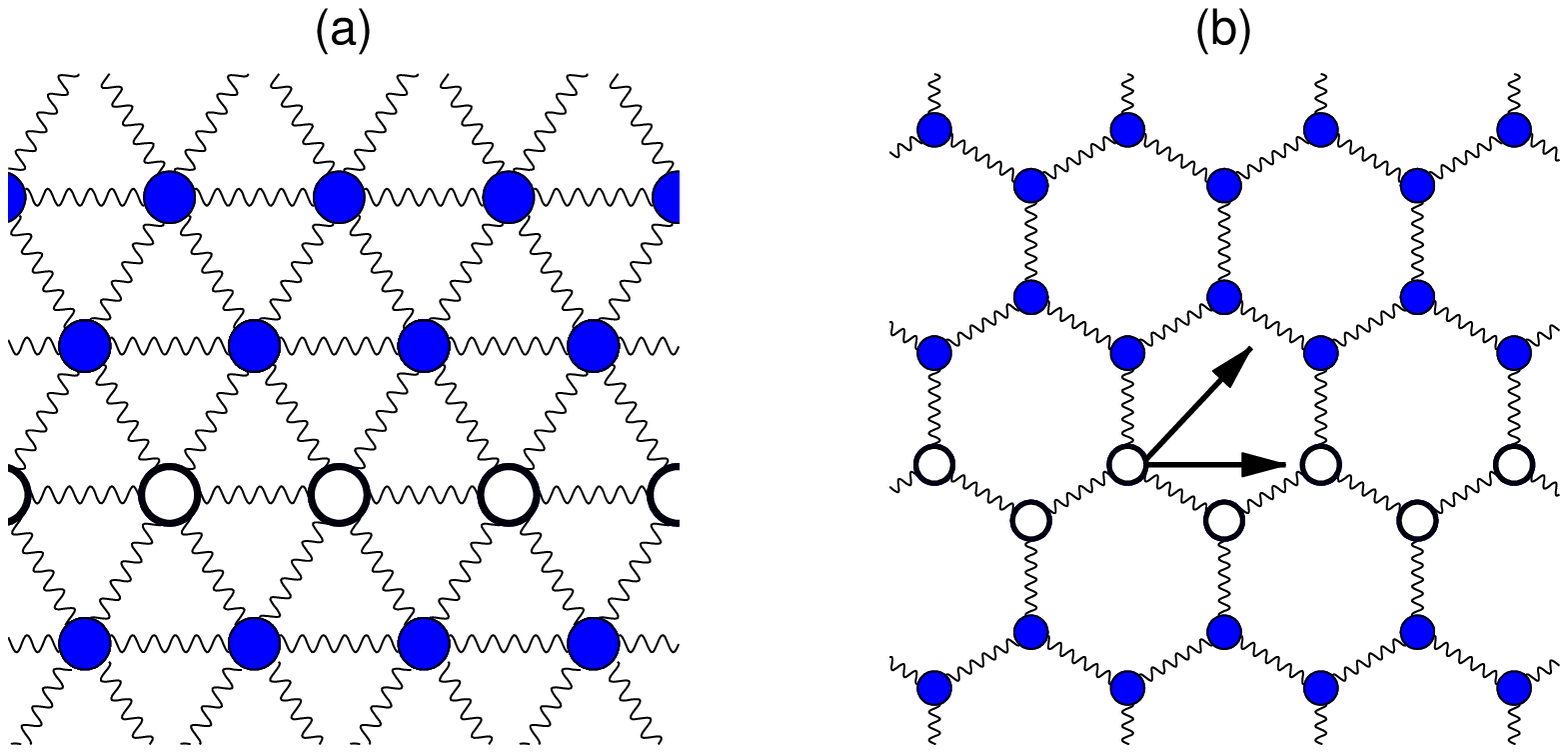}
\end{center}
\caption{\small {The line defect in the otherwise perfect lattices are shown for the hexagonal (a) and honeycomb structure (zigzag defect) (b), where the mass defects are indicated by the brightly shaded circles with a broad outline.}}
\label{fig:defect_lattice}
\end{figure}
\vspace{0.2cm}
\subsection{Hexagonal Lattice}
We initially consider the lattice shown in figure \ref{fig:defect_lattice}(a), where the line defect located at $m=0$ contains masses with values ($M_0$) which differ from those in the bulk lattice ($M_1$). In accordance with this alteration to the perfect structure, we adjust the difference equation \eqref{eq:triangular_difference} to give
\vspace{-0.4cm}
\beq
y_{n+1,m}+y_{n-1,m}+y_{n,m+1}+y_{n,m-1}+y_{n-1,m+1}+y_{n+1,m-1}-6 y_{n,m}=-\Omega^2 y_{n,m}\begin{cases}M_1 \hspace{0.5cm} m \neq0 \cr
   M_0  \hspace{0.5cm} m=0 \end{cases}.
\label{eq:triangular_difference_defect} 
\eeq For the sake of brevity we shall apply our multiple-scale scheme purely in the direction of the line defect, where the short-scale is now characterised by  the discrete variable $N=(0, \pm 1)$; this represents a diagonal-column of masses in the direction of the ${\bf e}_2$ axis (figure \ref{fig:triangle_dispersion}(a)) and its two nearest neighbouring columns. It is worth noting that if we were to apply our asymptotic expansion in both $(\eta_1,\eta_2)$ directions we would be limited to observing the oscillations associated to $\mathcal{O}(\epsilon^2)$ mass variations, this is unlike the discrete $m$ case where,in addition, we are able to consider the case $M_0\ll 1$.
\\

\begin{figure}
\begin{center}
    \includegraphics[height=5cm, width=10cm]{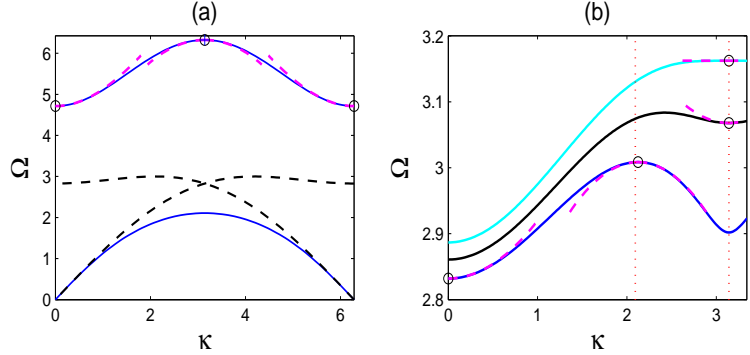}
\end{center}
\caption{\small {Panel (a) shows the dispersion curve in the range for $M_1=M_0=1$ with a set of dashed curves and $M_1=1, M_0=0.4$ is shown with the solid lines. The defect-free dispersion curves are identical to the curve in figure \ref{fig:triangle_dispersion}(c). The dashed lines, which are asymptotes to the solid curve, are deduced from equation \eqref{eq:tri_defect_freq_correction}. Panel (b) demonstrates the moving standing wave frequency where we once again take $M_1=1$. The solid curves have mass values $M_0=0.95$(bottom), $0.85$ (middle) and $0.8$  (upper) . The asymptotes derived using our method are shown for the solid curve ($M_0=0.95$) at $\kappa=0$ and at the location of the moving standing wave using our two-scale approach. Whilst the asymptotics at $\kappa=\pi$ for the curves $M_0=0.85, 0.8$ are deduced from the integral \eqref{eq:tri_defect_integral}. The dotted vertical lines represent the initial and final position of the moving standing wave, $2\pi/3$ and $\pi$ respectively. }}
\label{fig:tri_defect_disp}
\end{figure}

As was discussed earlier we opt to expand the macroscale using solely the horizontal coordinate in the rectangular lattice system, $\eta=\epsilon(n+m/2)$. Hence the displacement $y_{n,m}$ is denoted by
\beq
y_{n,m}=y_m\left(N,\eta \right),
\label{eq:tri_defect_displacement}\eeq  where in our macroscale the line defect will be interpreted as a continuum interface separating two structures which will have a hexagonal geometry implicitly defined.  We assume a constant phase shift $\kappa$ between two neighbouring diagonal columns of masses, where  $\kappa$ is defined in the $\eta$ direction and takes values in the range $[0, 2\pi)$. An example of a neighbouring mass displacement to $y_{n,m}$,  written in the two-scale notation is 
\beq
y_{n+1,m-1}=y_{m-1}\left(1,\eta+ \frac{\epsilon}{2} \right)=\exp\left(i \kappa \right) y_{m-1}\left(0,\eta+ \frac{\epsilon}{2} \right).
\label{eq:tri_defect_expansion_example}\eeq Using this we eventually derive the two-scale extension of the difference equation \eqref{eq:triangular_difference_defect}
\beq
e^{i \kappa}\left[y_{m}(\eta+\epsilon)+y_{m-1}\left(\eta+\frac{\epsilon}{2}\right)\right]+ e^{-i \kappa}\left[y_{m}(\eta-\epsilon)+y_{m+1}\left(\eta-\frac{\epsilon}{2}\right)\right]+ y_{m-1}\left(\eta-\frac{\epsilon}{2} \right)
\nonumber\eeq
\vspace{-0.2cm}
\beq
+y_{m+1}\left(\eta+\frac{\epsilon}{2} \right)-6y_m(\eta)=-M_1\Omega^2 y_m\left(\eta \right)-\Omega^2\left(M_0-M_1 \right)y_m\left(\eta \right) \delta_{m,0},
\label{eq:tri_defect_longscale_eqn}\eeq
 where we have suppressed the short-scale coordinate such that $y_m(0,\eta)=y_m(\eta), \forall  m \in \mathbb{Z}, \eta \in \mathbb{R}$. The above displacement functions are Taylor expanded for small $\epsilon$, in a similar manner to equation \eqref{eq:triangle_taylors}, albeit in a single direction $\eta$ in place of $(\eta_1,\eta_2)$. Subsequently we adopt the expansion shown in equation \eqref{eq:triangular_displ_expansion} for $\Omega^2$ and an analogous ansatz for the displacement function is used,
 \beq
 y_m(\eta)=y^{(0)}_m(\eta)+\epsilon y^{(1)}_m(\eta)+\epsilon^2 y^{(2)}_m(\eta)+\mathcal{O}\left(\epsilon^3\right)
\label{eq:tri_defect_displacement_ansatz} \eeq This expansion is substituted into equation \eqref{eq:tri_defect_longscale_eqn} to give us the following leading order problem
 \beq
 e^{i \kappa}\left[y^{(0)}_{m}+y^{(0)}_{m-1}\right]+ e^{-i \kappa}\left[y^{(0)}_{m}+y^{(0)}_{m+1}\right]+ y^{(0)}_{m-1}+y^{(0)}_{m+1}-6y^{(0)}_m=-M_1\Omega_0^2 y^{(0)}_m-\Omega_0^2\left(M_0-M_1 \right)y^{(0)}_m \delta_{m,0},
\label{eq:leading_order_tri_defect}\eeq where notationally here and hereafter we shall use the convention $y^{(j)}_m=y^{(j)}_m(\eta)$. Note that in the above semi-discrete equation there is no explicit dependence on $\eta$ hence $y^{(0)}_m=f(\eta)Y^{(0)}_m$, where the function $f(\eta)$ describes the envelope modulation in the direction of the defect. The leading order problem \eqref{eq:leading_order_tri_defect} is solved using the one-dimensional form of the Fourier transform defined in equations \eqref{eq:forwardDFT} and \eqref{eq:backwardDFT},
\beq
\tilde{Y}^{(j)}(\alpha)=\sum_{m} Y^{(j)}_{m} \exp\left(-m\frac{i}{2} \left[\kappa + \sqrt{3}\alpha \right] \right), \hspace{0.2cm} Y^{(j)}_{m}=\frac{\sqrt{3}}{4\pi} \int\limits_{-2\pi/\sqrt{3}}^{2\pi/\sqrt{3}}  \tilde{Y}^{(j)}(\alpha) \exp\left(m\frac{i}{2} \left[\kappa + \sqrt{3}\alpha \right] \right) d\alpha.
\label{eq:tri_defect_FT} \eeq Subsequently after applying the above transform to the leading order problem and resolving the ensuing equation for $Y_m^{(0)}$ we obtain the following integral
\beq
Y_m^{(0)}=\frac{\Omega_0^2}{2\pi}\int\limits_{-\pi}^{\pi}\frac{Y_0^{(0)}\left(M_1-M_0 \right)e^{im(\kappa +2\beta)/2}}{4\cos(\kappa/2)\cos(\beta)+2\cos(\kappa)+\Omega_0^2M_1-6} d\beta 
\label{eq:tri_defect_integral}\eeq We integrate the above to derive the displacement in $m$,
\beq 
Y^{(0)}_m=Y^{(0)}_0 \exp \left(\frac{i}{2}m\kappa \right) \left[-\frac{\Omega_0^2(M_0-M_1)-(2\cos(\kappa)+\Omega_0^2M_1-6)}{4\cos(\kappa/2)} \right]^{|m|} \hspace{0.2cm} (\kappa \neq \pi)
\label{eq:leading_order_tri_defect_disp}\eeq and for $\kappa=\pi, Y^{(0)}_m \neq 0 \Leftrightarrow m=0$, additionally the leading order frequency term is found by solving the following dispersion relation,
\beq
\sqrt{(2\cos(\kappa)+\Omega_0^2M_1-6)^2-16\cos^2\left(\frac{\kappa}{2}\right)}=-\Omega_0^2(M_0-M_1).
\label{eq:tri_defect_dispersion_relation}\eeq 
\vspace{0.1cm} Note that the Bloch solution valid precisely at the standing wave frequency is identical to the leading order asymptotic solutions deduced above. The exact solution of the dispersion relation, $\Omega^2$, is shown for different $M_0$ values in figures \ref{fig:tri_defect_disp}(a) and (b).
\\

Returning to the asymptotics recall that our interest is in analysing the asymptotes about those standing wave frequencies in figure \ref{fig:tri_defect_disp} which display quadratic behaviour, hence for these frequencies $\Omega_1=0$. We proceed to $\mathcal{O}(\epsilon^2)$ where after Fourier transforming the second-order governing equation we derive the following 
\beq
\tilde{Y}^{(2)}(\alpha)f^{(2)}(\eta)+\frac{(M_0-M_1)\Omega_0^2y_{0}^{(2)}(\eta)}{4\cos(\kappa/2)\cos(\sqrt{3}/2\alpha)+2\cos(\kappa)+\Omega_0^2M-6}=
\nonumber\eeq
\beq
\frac{Y_m^{(0)} \left[-\frac{1}{2}\cos(\kappa/2)\cos(\sqrt{3}/2\alpha)f_{\eta,\eta}-\cos(\kappa)f_{\eta,\eta}-\Omega_2^2M_1f(\eta)\right]+\Omega_2^2(M_1-M_0)f(\eta)Y^{(0)}_0}{4\cos(\kappa/2)\cos(\sqrt{3}/2\alpha)+2\cos(\kappa)+\Omega_0^2M-6},
\label{eq:tri_defect_oreeps2_eqn}\eeq where $f_{\eta,\eta}=d^2f(\eta)/d\eta^2$. The inverse Fourier transform is applied to the above equation and the Fredholm solvability condition is invoked,  this enables us to derive an ODE dependent explicitly on the long-scale variable $\eta$. In order to keep the algebra succinct, only the ODE's for fixed $\kappa=0,\pi$ (periodic and anti-periodic oscillations along the $\eta$ direction) are shown

\beq
\frac{M_1\Omega_0^2-6}{4M_1}f_{\eta,\eta}+\Omega_2^2f=0 \hspace{0.2cm}(\kappa=0),\hspace{0.5cm}f_{\eta,\eta}-\frac{8}{\Omega_0^2}\Omega_2^2f=0\hspace{0.2cm}(\kappa=\pi).
\label{eq:tri_defect_ODE}\eeq To get the asymptotics we apply the Bloch condition to the envelope function $f(\eta)$ such that $f(\eta+\epsilon)=\exp[i(\kappa-\psi)]f(\eta)$ ($\psi$ is location of the standing wave), by assuming a solution $f(\eta)=\exp(i\eta)$ it follows that $\epsilon=(\kappa-\psi)$; using this property we derive the second-order frequency corrections for $\kappa=0, \pi$ as 
\beq
\Omega^2=\frac{8M_1}{M_0(2M_1-M_0)}+\kappa^2\frac{M_1 \Omega_0^2-6}{4M_1}+\mathcal{O}[\kappa^3], \hspace{0.4cm}\Omega^2=\frac{8}{M_0}-(\kappa-\pi)^2\frac{\Omega_0^2}{8}+\mathcal{O}[(\kappa-\pi)^3].
\label{eq:tri_defect_freq_correction}\eeq Note that these solutions pertain to the optical branch of the dispersion curve and are shown plotted in figure \ref{fig:tri_defect_disp}(a) and (b).
\\

\begin{figure}
\begin{center}
    \includegraphics[height=5cm, width=10cm]{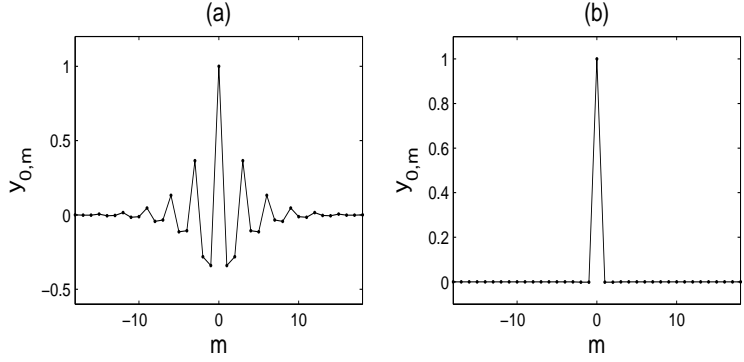}
\end{center}
\caption{\small { This figure demonstrates the decay of the displacement along the ${\bf e}_2$ axis of the hexagonal lattice. In panel (a) $M=1,M_0=0.93, \kappa \approx 2.153$, whilst in (b),$M=1,M_0=0.75, \kappa=\pi$. }}
\label{fig:tri_defect_displacements}
\end{figure}

 The dashed curves ($M_1=M_0$) shown in figure \ref{fig:tri_defect_disp}(a) illustrate the derivation of the highest branch of our line-defect dispersion curves. The selected curves plotted along $\kappa=0\rightarrow\pi$, for the defect-free lattice, are also present in the figure \ref{fig:triangle_dispersion}(c) along the path ACB in the irreducible zone. Traversing the Brillouin zone in this direction corresponds to $\kappa_2=0$ and the reverse dashed curve shown in figure \ref{fig:tri_defect_disp}(a)  is found at $\kappa_2=\pm 2\pi/\sqrt{3}$, $\kappa=\kappa_1:0\rightarrow2\pi$. As a result, if we were to set $M_1=M_0$ in the $\kappa=0$ ODE \eqref{eq:tri_defect_ODE}, it would resemble the $\eta_2$ independent version of the hyperbolic equation \eqref{eq:triangle_pde}.  An additional observation is that previously the values of $\kappa$ which represented standing wave frequencies for the perfect structure  in the range $[0,\pi]$ were located at $0$ and $2\pi/3$, however  with the introduction of the line defect, the standing wave initially located at $2\pi/3$, shifts to $\pi$ as the value of $M_0$ decreases. This phenomenon is due to the fact that when  $M_0 \ll 1$ the wave is almost entirely localised in the $\eta$-direction ($m=0$) hence we are dealing with a quasi-one-dimensional problem as shown in figure \ref{fig:tri_defect_displacements}(b). This results in a distortion of the original hexagonal Brillouin zone into a single line $\kappa:0\rightarrow \pi$ where standing waves are present at either end of the new irreducible zone.
\\

A disadvantage of our quasi-one-dimensional two-scale approach is related to the notable exclusion of the cross derivative $\partial^2/\partial\eta_1\partial\eta_2$ in our expansion. The presence of which is attributed to the non-orthogonal geometry under consideration. It follows that the contribution of this term to the two-dimensional expansion, used for the perfect hexagonal lattice, dictates the accuracy of our single variable asymptotic scheme. The coefficient of this cross derivative in the two-dimensional asymptotic scheme is $C \sin(\kappa_1/2)\sin\left(\kappa_2\sqrt{3}/2\right)$ ($C$ is a constant), hence for $\kappa=0$ our scheme is accurate $\forall M_0 \leq M$. For $\kappa=\pi$ the asymptotic method is precise in describing the local frequency variation solely for the case $M_0 \ll 1$. This is due to the absence of the cross derivative as the wave is completely localised along the defect, figure \ref{fig:tri_defect_disp}(a).
\\

In order to derive a uniformly valid frequency correction for the moving standing wave frequency and at $\kappa=\pi$, we asymptotically expand both $\kappa$ and $\Omega_0$, in powers of $\epsilon$, in the integrand of equation \eqref{eq:tri_defect_integral} (set $m=0$) and solve the ensuing equations in $\mathcal{O}(\epsilon)$. Therefore by using this approach we find that the asymptotic frequency, valid for all $M_0$ values, at $\kappa=\pi$ is 
\beq
\Omega^2=\frac{8}{M_0}-(\kappa-\pi)^2\frac{\Omega_0^2 \left(M_1 \Omega_0^2-10 \right)}{8(M_1\Omega_0^2-8)}+\mathcal{O}[(\kappa-\pi)^3].
\label{eq:backwards_pi_freq}
\eeq 
When $M_0 \ll 1$ it is seen that $\Omega_2^2$ in the above equation converges to the frequency correction in equation \eqref{eq:tri_defect_freq_correction}. Additionally we could easily deduce the accompanying ODE, governing motion about $\kappa=\pi$, by working backwards from the solution \eqref{eq:backwards_pi_freq}.  


\vspace{0.1cm}

\subsection{Honeycomb Lattice}
In this section we shall introduce two distinct types of infinite line-defects into the honeycomb structure, namely the zigzag and armchair defects. Interestingly we will have to adjust our two-scale approach  now that the difference equations are coupled. 

\vspace{0.15cm}
\subsubsection{Zigzag defect}
 Initially we consider the zigzag defect, illustrated in figure \ref{fig:defect_lattice}(b). The defective mass value is once again denoted by $M_0$ whilst the remaining mass values are given by $M_1$. The altered equations of motion are 
\beq
q_{n,m-1}+q_{n-1,m}+q_{n,m}-3p_{n,m}=-M_1\Omega^2 p_{n,m}-\left[M_0-M_1 \right]\Omega^2 p_{n,m}\delta_{m,0},
\label{eq:zigzag_difference_equations1} \eeq
\beq
p_{n,m}+p_{n,m+1}+p_{n+1,m}-3q_{n,m}=-M_1\Omega^2 q_{n,m}-\left[M_0-M_1 \right]\Omega^2 q_{n,m}\delta_{m,0}.
\label{eq:zigzag_difference_equations2} \eeq For convenience we rotate the lattice diagram, figure \ref{fig:honeycomb_dispersion}(a), by $\pi/3$ in the counter clockwise, whereby the ${\bf e}_1$ axis is now aligned with the horizontal axis. The relationship between the long-scale orthogonal coordinates and the newly defined basis system now resembles that of the hexagonal lattice \eqref{eq:triangle_lattice_transformation}, enabling us to use many ideas from that geometry.
\\

We opt to apply our homogenisation method in a single direction, where the short-scale is  characterised by $N=(0,\pm 1)$, which represents a diagonal column of masses in the direction of the ${\bf e}_2$ axis and its two nearest neighbours. Recall that the arrangement of masses in the honeycomb lattice is obtained by considering the hexagonal geometry, where each point in the hexagonal lattice consists of two basis masses. Hence comparatively, the aforementioned diagonal column of masses is comprised of  a single series of $p_{\hat{n},m}$ masses and another single series $q_{\hat{n},m}$, as opposed to a single set of  $y_{\hat{n},m}$ masses as was the case for the hexagonal lattice ($\hat{n}$ is a fixed integer value). The macroscale is once again taken as the horizontal component of the orthogonal system, therefore the displacement functions $p_{n,m},q_{n,m}$ take the form \eqref{eq:tri_defect_displacement}, whilst the detailing of the short-scale is identical to that shown in equation \eqref{eq:tri_defect_expansion_example}. We proceed in a similar manner to the hexagonal lattice, whereby we substitute in $p_{n+N,m+M}=\exp(i\kappa N)p_{m+M}(\eta + \epsilon[N+M/2])$ and where $q_{n+N,m+M}$ takes a similar form into the difference equations, \eqref{eq:zigzag_difference_equations1} and \eqref{eq:zigzag_difference_equations2}, ($N,M \in \mathbb{Z}$) . Thereafter we Taylor expand for small $\epsilon$ in the long-scale and apply the natural separation of scales to the displacement functions, \eqref{eq:tri_defect_displacement_ansatz} and the frequency \eqref{eq:triangular_displ_expansion}.
\\

\begin{figure}
\begin{center}
    \includegraphics[height=8cm, width=9cm]{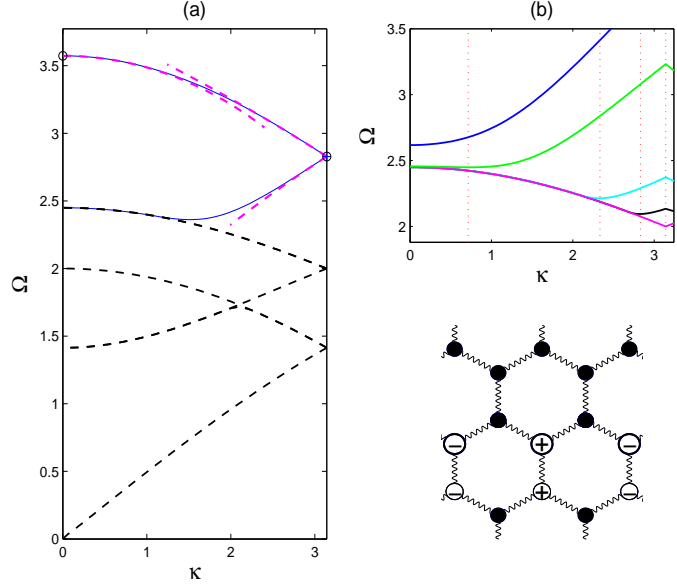}
\end{center}
\caption{\small {In panel (a) the mass value for the bulk lattice is $M_1=1$. The dashed lines are derived by travelling along the paths  ACB and its reverse in the irreducible Brillouin zone, where $M_0=M_1$. This section of the defect-free curve is also shown in figure \ref{fig:honeycomb_dispersion}(b).  The top two branches for the dispersion curve related to the zigzag defect lattice $M_0=0.4$ are shown as the solid lines. The asymptotes about the standing wave frequencies $\kappa=0,\pi$ are also shown as dashed lines. The upper Bloch diagram in panel (b) visually demonstrates the moving standing wave (second branch) for  $\kappa=\pi \rightarrow 0$ as $M_0=1 \rightarrow 0$ ($M_0=0.8,0.6,0.3,0.2$ are the values taken above). The lower lattice figure demonstrates the standing wave pattern at $\kappa=\pi$, where the relatively larger circles, with $\pm$ signs, indicate a displacement with greater absolute value than the smaller counterparts. The masses consisting of a solid dark circle indicate a zero perturbation from the equilibrium position.}}
\label{fig:zigzag_disp_lattice}
\end{figure}

After completing all the outlined expansions we arrive at leading order where we obtain the following
\beq
p_{m}^{(0)}(\eta)+p_{m+1}^{(0)}(\eta)+e^{i\kappa}p_{m}^{(0)}(\eta)-3q_{m}^{(0)}(\eta)+\Omega_0^2 M_1 q_{m}^{(0)}(\eta)+\Omega_0^2[M_0-M_1]\delta_{m,0}q_{m}^{(0)}(\eta)=0,
\eeq
\beq
q_{m}^{(0)}(\eta)+q_{m-1}^{(0)}(\eta)+e^{-i\kappa}q_{m}^{(0)}(\eta)-3p_{m}^{(0)}(\eta)+\Omega_0^2 M_1 p_{m}^{(0)}(\eta)+\Omega_0^2[M_0-M_1]\delta_{m,0}p_{m}^{(0)}(\eta)=0,
\eeq 
 as there is no explicit dependence on $\eta$ we redefine the displacements as $p_m^{(0)}=f(\eta)P_m^{(0)}$ and $q_m^{(0)}=f(\eta)Q_m^{(0)}$. Note that the elementary cell for the honeycomb lattice is defined as a single mass, associated to $p_{n,m}$, and its neighbouring mass, associated to $q_{n,m}$, hence it follows that the long-scale is only concerned with a single $\eta$-dependent function, $f(\eta)$.  We apply the semi-discrete Fourier transform \eqref{eq:tri_defect_FT} to the discrete components of the displacements in the above coupled equations,
\beq
\hspace{-0.5cm}\left(1+\exp\left(i\kappa\right)+\exp\left[i\left(\kappa +\sqrt{3}\alpha\right)/2\right] \right)\tilde{P}^{(0)}(\alpha) f(\eta)+[M_1 \Omega_0^2-3]\tilde{Q}^{(0)}(\alpha) f(\eta)+\Omega_0^2 q_0^{(0)}(\eta)(M_0-M_1)=0,
\label{eq:hon_def_FT_eqn1} \eeq
\beq
\hspace{-0.5cm}\left(1+\exp\left(-i\kappa\right)+\exp\left[-i\left(\kappa +\sqrt{3}\alpha\right)/2\right] \right)\tilde{Q}^{(0)}(\alpha) f(\eta)+[M_1 \Omega_0^2-3]\tilde{P}^{(0)}(\alpha) f(\eta)+\Omega_0^2 p_0^{(0)}(\eta)(M_0-M_1)=0.
\label{eq:hon_def_FT_eqn2} \eeq The above equations are resolved for  $\tilde{P}^{(0)}(\alpha)$ and $\tilde{Q}^{(0)}(\alpha)$ and then inverse Fourier transformed \eqref{eq:tri_defect_FT},
\beq
\hspace{-1cm} P_m^{(0)}=\left(M_1-M_0\right)\frac{\sqrt{3}}{4\pi^2}\int\limits_{-2\pi/\sqrt{3}}^{2\pi/\sqrt{3}} \frac{\Omega_0^2 \left[P_0^{(0)}\left(e^{i \left[\kappa + \sqrt{3}\alpha \right]/2 }+e^{i\kappa}+1 \right) - Q_0^{(0)}\left(M_1\Omega_0^2-3 \right) \right]e^{\left(i m\left[\kappa + \sqrt{3}\alpha \right]/2 \right)}}{\left[4\cos\left(\kappa/2 \right)\cos\left(\sqrt{3}\alpha/2\right)+2\cos\left(\kappa\right)-6-M_1^2\Omega_0^4+6M_1\Omega_0^2 \right]} d\alpha, \label{eq:hon_defect_P_IFT_eqn}\eeq 
\beq
\hspace{-1cm} Q_m^{(0)}=\left(M_1-M_0\right)\frac{\sqrt{3}}{4\pi^2}\int\limits_{-2\pi/\sqrt{3}}^{2\pi/\sqrt{3}} \frac{\Omega_0^2 \left[Q_0^{(0)}\left(e^{-i \left[\kappa + \sqrt{3}\alpha \right]/2 }+e^{-i\kappa}+1 \right) - P_0^{(0)}\left(M_1\Omega_0^2-3 \right) \right]e^{\left(i m\left[\kappa + \sqrt{3}\alpha \right]/2 \right)}}{\left[4\cos\left(\kappa/2 \right)\cos\left(\sqrt{3}\alpha/2\right)+2\cos\left(\kappa\right)-6-M_1^2\Omega_0^4+6M_1\Omega_0^2 \right]} d\alpha.
\label{eq:hon_defect_Q_IFT_eqn}\eeq 
The dispersion relation, relating $\Omega_0$ and $\kappa$, is found by setting $m=0$ in the above equations and solving them simultaneously. We have omitted showing a more explicit representation of $\Omega_0(\kappa)$ due to the amount of algebra present in the solution however the explicit dispersion relation is shown visually in figure \ref{fig:zigzag_disp_lattice}. 


\vspace{0.5cm}

\indent As we have implemented a high-frequency homogenisation methodology our primary focus will be on the highest branch of the dispersion diagram (figure \ref{fig:zigzag_disp_lattice}(a)), which is spawned from the two underlying curves representing the defect-free lattice ($M_1=M_0$), see figure \ref{fig:zigzag_disp_lattice}(a). These two curves are derived from travelling along the path ACBD and its reverse in the hexagonal Brillouin zone (figure \ref{fig:triangle_dispersion}(b)), as a result the lattice containing a zigzag defect has an inherent symmetry about $\kappa=\pi$. Additionally the introduction of a defect results in a new standing wave mode at $\kappa=\pi$, however unlike the hexagonal lattice case there is no additional standing wave within the range $0<\kappa<\pi$ for the highest branch, as is demonstrated in figure \ref{fig:zigzag_disp_lattice}(a). However there is a moving standing wave present along the second highest branch of the zigzag defect dispersion curve \ref{fig:zigzag_disp_lattice}(b), which disappears at $M_0=0.2$ and travels from $\kappa=\pi \rightarrow 0$ as $M_0=1 \rightarrow 0$. Interestingly at the new standing wave $\kappa=\pi$ the oscillation pattern is such that $P^{(0)}_\gamma, Q^{(0)}_\theta \neq 0 \Leftrightarrow \gamma=0,\theta=-1$ (or by symmetry $\gamma=1,\theta=0$) , where the displacements are primarily along the zigzag defect, for $M_0=0.4, M_1=1$, $P^{(0)}_0=-1, Q^{(0)}_{-1}=0.2$, figure \ref{fig:zigzag_disp_lattice}(b).
\\

We now return to detailing our asymptotic method by illustrating the methodology at $\kappa=0$. Note that we have omitted extraneous algebra when it becomes substantial in volume by providing numerical solutions for fixed parameter values.  Dissimilar to the hexagonal case we are now dealing with a two-dimensional lattice system containing a row of defective masses hence the two scale approach will be slightly adjusted from the method used in the previous section.  At leading order we rearrange the coupled integral equations \eqref{eq:hon_defect_P_IFT_eqn}, \eqref{eq:hon_defect_Q_IFT_eqn} ($m=0$) into the following matrix equation
\beq
\left[A^{(0)}\left(\Omega_0\right)\right]{\bf P_0^{(0)}}=0, \hspace{0.3cm} {\bf P_0^{(0)}}=\left[P_0^{(0)}, Q_0^{(0)} \right]^T, \hspace{0.4cm} \left[A^{(0)}\left(\Omega_0\right)\right]_{1,1}= 1+\frac{\Omega_0^2\left(M_0-M_1 \right) \left(M_1\Omega_0^2-3 \right)}{\sqrt{-16+\left(4-6\Omega_0^2M_1+\Omega_0^4M_1^2 \right)}},
\label{eq:honeycomb_leading_order}
\eeq
\beq
\hspace{-0.4cm}
\left[A^{(0)}\left(\Omega_0\right)\right]_{1,2}=\frac{\Omega_0^2\left(M_0-M_1 \right)\left[-8+\sqrt{-16+\left(4-6\Omega_0^2M_1+\Omega_0^4M_1^2 \right)}-\left(4-6\Omega_0^2M_1+\Omega_0^4M_1^2 \right) \right]}{4\sqrt{-16+\left(4-6\Omega_0^2M_1+\Omega_0^4M_1^2 \right)}},
\nonumber\eeq where $\left[A^{(0)}\left(\Omega_0\right)\right]_{i,j}$ denotes the $(i,j)$'th component and
$A^{(0)}\left(\Omega_0\right)$ is a $2$x$2$ symmetric matrix with $\left[A^{(0)}\left(\Omega_0\right)\right]_{1,1}=\left[A^{(0)}\left(\Omega_0\right)\right]_{2,2}$. Thereafter we derive the leading order frequency where for the case $M_1=1, M_0=0.4$ we obtain $\Omega_0=\pm 3.573134127$ and  $\pm 1.915456675$. After substituting in the aforementioned eigenvalues we derive an eigenvector which in turn provides the relation $P_0^{(0)}=-Q_0^{(0)}$.
\\

We now move onto $\mathcal{O}(\epsilon)$, where after Fourier transforming the initial equation we obtain 
\beq
\hspace{-0.6cm}
\left(P_m^{(0)}f_{\eta}/2+p_m^{(1)} \right)\left(e^{i\sqrt{3}\alpha/2}+2\right)+q_m^{(1)}\left(M_1\Omega_0^2-3 \right)+\Omega_1^2 M_1 Q_m^{(0)}f+\left(\Omega_0^2 q_0^{(1)}+\Omega_1^2 Q_0^{(0)}f \right)\left(M_0-M_1 \right)=0,
\label{eq:honeycomb_ordeps_FT_P}
\eeq
\beq
\hspace{-0.8cm}
\left(-Q_m^{(0)}f_{\eta}/2+q_m^{(1)} \right)\left(e^{-i\sqrt{3}\alpha/2}+2\right)+p_m^{(1)}\left(M_1\Omega_0^2-3 \right)+\Omega_1^2 M_1 P_m^{(0)}f+\left(\Omega_0^2 p_0^{(1)}+\Omega_1^2 P_0^{(0)}f \right)\left(M_0-M_1 \right)=0.
\label{eq:honeycomb_ordeps_FT_Q}
\eeq We perform a series of substitutions involving the above equations through which we  deduce two subsequent formulas for $p_m^{(1)}$ and $q_m^{(1)}$, where each formula is independent of the opposing $\mathcal{O}(\epsilon)$ displacement function. These formulae take a form similar to that of equation \eqref{eq:tri_defect_oreeps2_eqn} and therefore are rearranged into the matrix equation
\beq
\left[A^{(0)}\left(\Omega_0\right)\right]{\bf P_0^{(1)}}=\left[A^{(1)}\left(\Omega_1\right)\right]{\bf P_0^{(0)}}, \hspace{0.2cm} \left[A^{(j)}\left(\Omega_j\right)\right] \hspace{0.1cm}\text{symmetric for } \hspace{0.1cm} j \in \mathbb{Z}_{\geq 0},
\label{eq:honeycomb_ordeps_matrix}\eeq Using the knowledge that $\left[A^{(0)}\left(\Omega_0\right)\right]$ is self-adjoint we deduce the solvability condition ${\bf \left[P_0^{(0)}\right]}^T \left[A^{(1)}\left(\Omega_1\right)\right]{\bf P_0^{(0)}}=0$ which gives us the relations $p_0^{(1)}=Q_0^{(0)} f_{\eta}/2-q_0^{(1)}$ and $\Omega_1=0$.
\\

Finally at $\mathcal{O}(\epsilon^2)$ we Fourier transform the governing equations thereby giving us a set of equations which resemble \eqref{eq:honeycomb_ordeps_FT_P}, \eqref{eq:honeycomb_ordeps_FT_Q} albeit with the displacement terms of a higher order, the addition of an $\Omega_2^2$ term and a second-derivative of $f(\eta)$. After some algebra we once again deduce an eigenvalue problem
\beq
\left[A^{(0)}\left(\Omega_0\right)\right]{\bf P_0^{(2)}}=\left[A^{(1)}\left(\Omega_1\right)\right]{\bf P_0^{(1)}}+\left[A^{(2)}\left(\Omega_2\right)\right]{\bf P_0^{(2)}},
\label{eq:honeycomb_ordeps2_matrix}\eeq where the presence of the first-order eigenvector, ${\bf P_0^{(1)}}$, is attributed to the inhomogeneity  of the first-order matrix equation after prescribing the $\Omega_0$ value. We apply a similar solvability condition, that was utilised at first-order, whereby we multiply the right-hand side of equation \eqref{eq:honeycomb_ordeps2_matrix} by $\left[{\bf P_0^{(0)}}\right]^T$ thereby giving us an ODE of the form
$f_{\eta,\eta}+\tau \Omega_2^2f=0$. For $M_1=1, M_0=0.4$ we obtain $\tau=(0.6097781588)^{-1}$. In order to derive the asymptote about $\kappa=0$ we apply the Bloch periodicity condition $f(\eta)=\exp\left(i\kappa\eta \right)$ to the prior ODE, the resulting asymptote is shown in figure \ref{fig:zigzag_disp_lattice}(a).
\\

Note that a similar methodology is implemented when deriving the governing ODE about  $\kappa=\pi$. For the case $M_1=1, M_0=0.4$ we deduce the eigenvalues $\Omega_0=\pm 2\sqrt{2},\pm 1.118033989\sqrt{2} $ where after substitution into $\left[A^{(0)}\left(\Omega_0\right)\right]$ we obtain the zero matrix. Hence the leading order solution takes the form ${\bf P_0^{(0)}}=f^{(1)}(\eta){\bf i}+f^{(2)}(\eta){\bf j}$, where ${\bf i}, {\bf j}$ are the unit orthogonal vectors. At $\mathcal{O}(\epsilon)$ after forward and inverse Fourier transformations we find that $\left[A^{(1)}\left(\Omega_1\right)\right]{\bf P_0^{(0)}}={\bf 0}$ which gives us the following coupled system
\beq
f^{(2)}_{\eta}+\tau\Omega_1^2 f^{(1)}=0, \hspace{0.2cm} f^{(1)}_{\eta}-\tau\Omega_1^2 f^{(2)}=0, \hspace{0.3cm}\tau=0.6633249584.
\label{eq:hon_defect_pi_ode} \eeq  An important point to note is that unlike the hexagonal line-defect case, the ODE's  found at $\kappa=0, \pi$ are uniformly accurate for all values of $M_0$. For the $\kappa=0$ case this is justified by observing that the mixed derivative is conspicuously absent for all $\kappa_2$ in the two-variable expansion whilst for $\kappa=\pi$ the same is true due to the local variation being linear. 
\\

An additional point to note is that we could have equivalently found the frequency corrections by asymptotically expanding about the desired wavenumber, frequency and associated displacement functions in the integral equations \eqref{eq:hon_defect_P_IFT_eqn}, \eqref{eq:hon_defect_Q_IFT_eqn} and solving accordingly for $\Omega_1,\Omega_2$. 



\subsubsection{Armchair defect}
\begin{figure}[htb!]
\begin{center}
    \includegraphics[height=6cm, width=8cm]{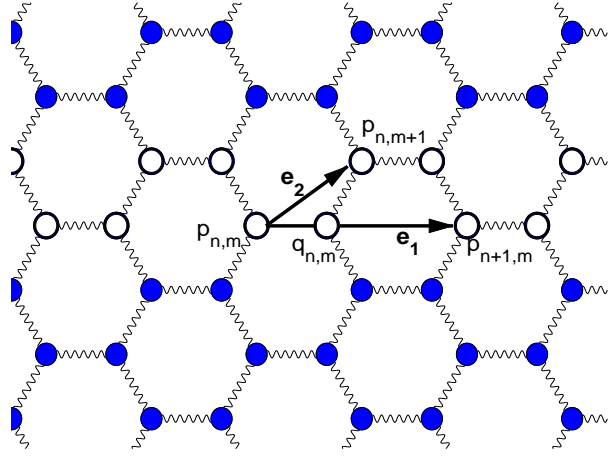}
\end{center}
\caption{\small {The armchair defect in the otherwise perfect lattices is shown for the honeycomb structure. The mass defects are indicated by the brightly shaded circles with a broad outline and the lattice basis vectors are labelled ${\bf e}_1$ and ${\bf e}_2$.}}
\label{fig:armchair_lattice}
\end{figure}

\begin{figure}[htb!]
\begin{center}
    \includegraphics[height=8.75cm, width=11cm]{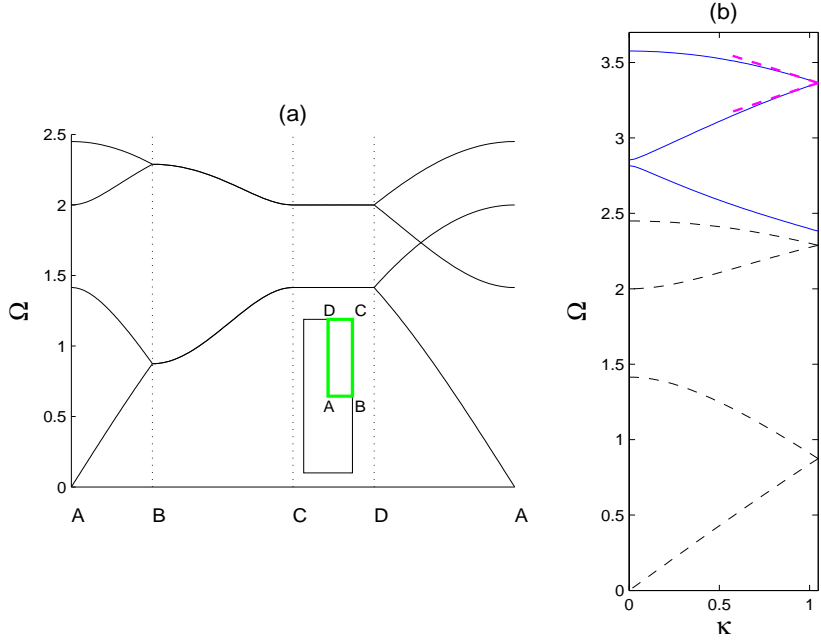}
\end{center}
\caption{\small {Panel (a) shows the dispersion curve $\Omega(\kappa_1,\kappa_2)$ for the perfect lattice where we have used the orthogonal formulation and plotted the dispersion relation along the edge of the irreducible Brillouin zone (the $ABCD$ rectangle in the inset). The top two solid curves in panel (b)are from the armchair defect lattice ($M_1=1, M_0=0.4$) where the asymptotic solution is shown by the dashed lines about $\kappa=\pi/3$ on the uppermost curve. The dashed lines below the solid curves represent the original dispersion curve ($M_1=M_0$). This latter curve is also present along $AB$ in (a). }}
\label{fig:armchair_2scale}
\end{figure}

An alternative defect pattern in the honeycomb lattice is that of the armchair defect, illustrated in figure \ref{fig:armchair_lattice}. Our axes of choice are ${\bf e}_1=3\epsilon{\bf i}, {\bf e}_2=\sqrt{3}\epsilon\left[\sqrt{3}/2{\bf i}+1/2{\bf j}\right]$ and our long-scale will be defined as $\eta_1=3\epsilon \left[n+m/2 \right], \eta_2=\epsilon \sqrt{3}/2 m$, it follows that the coupled equations of motion in the lattice are 
\beq
q_{n,m}+q_{n-1,m+1}+q_{n,m-1}-3 p_{n,m}=-M_1\Omega^2 p_{n,m}-\left[M_0-M_1 \right]\Omega^2 p_{n,m} \left(\delta_{m,0}+\delta_{m,1}\right),
\label{eq:armchair_eqn1}\eeq
\beq
p_{n,m}+p_{n,m+1}+p_{n+1,m-1}-3 q_{n,m}=-M_1\Omega^2 q_{n,m}-\left[M_0-M_1 \right]\Omega^2 q_{n,m} \left(\delta_{m,0}+\delta_{m,1} \right).
\label{eq:armchair_eqn2}\eeq The formulation of our two-scale method follows in a similar manner to the zigzag structure where the elementary cell contains two masses, the displacements of which are denoted by $p_{n,m}, q_{n,m}$ and are shown in figure \ref{fig:armchair_lattice}. 
The short-scale is characterised $N=(0,\pm 1)$, which is related to the masses in the cell and their nearest neighbours. We assume a constant phase shift between the columns of masses such that 
\beq
p_{n+1,m-1}=p_{m-1}(1,\eta_1+3\epsilon)=\exp \left(3 i \kappa \right)p_{m-1}\left(\eta_1+3\epsilon \right),
\label{eq:armchair_displacement_short_scale_expansion} \eeq and similarly for $q_{n+1,m-1}$.We substitute the above detailing of the short-scale oscillations into the difference equations \eqref{eq:armchair_eqn1}, \eqref{eq:armchair_eqn2}, for both $p_{n,m}$ and $q_{n,m}$  and persevere with our asymptotic method by Taylor expanding out the displacement functions and utilising the separation of scales to eventually derive the following leading order problem
\beq 
\hspace{-0.7cm}q_{m}^{(0)}(\eta)+q_{m-1}^{(0)}(\eta)+e^{-3i\kappa}q_{m+1}^{(0)}(\eta)-3p_{m}^{(0)}(\eta)+\Omega_0^2 M_1 p_{m}^{(0)}(\eta)+\Omega_0^2[M_0-M_1]\left(\delta_{m,0}+\delta_{m,1}\right)p_{m}^{(0)}(\eta)=0,
\label{eq:armchair_leadingorder1}\eeq 
\beq 
\hspace{-0.7cm}p_{m}^{(0)}(\eta)+p_{m+1}^{(0)}(\eta)+e^{3i\kappa}p_{m-1}^{(0)}(\eta)-3q_{m}^{(0)}(\eta)+\Omega_0^2 M_1 q_{m}^{(0)}(\eta)+\Omega_0^2[M_0-M_1]\left(\delta_{m,0}+\delta_{m,1}\right)q_{m}^{(0)}(\eta)=0,
\label{eq:armchair_leadingorder2}\eeq where hereafter $\eta=\eta_1$. It follows that $p_{m}^{(0)}=f(\eta)P_{m}^{(0)},q_{m}^{(0)}=f(\eta)Q_{m}^{(0)}$, where we now apply the forward component of the Fourier transform
\beq
\hspace{-0.7cm}\tilde{P}^{(j)}(\alpha)=\sum_{m} P^{(j)}_{m} \exp\left(-m\frac{i}{2} \left[3\kappa + \sqrt{3}\alpha \right] \right), \hspace{0.2cm} P^{(j)}_{m}=\frac{\sqrt{3}}{4\pi} \int\limits_{-2\pi/\sqrt{3}}^{2\pi/\sqrt{3}}  \tilde{P}^{(j)}(\alpha) \exp\left(m\frac{i}{2} \left[3\kappa + \sqrt{3}\alpha \right] \right) d\alpha,
\label{eq:armchair_defect_FT} \eeq and its counterpart relating to $Q^{(j)}_{m}$, to equations \eqref{eq:armchair_leadingorder1} and \eqref{eq:armchair_leadingorder2}. The ensuing equations are resolved for $\tilde{P}^{(0)}(\alpha), \tilde{Q}^{(0)}(\alpha)$ and inverse Fourier transformed to give the leading order displacements
\beq
\hspace{-1cm} P_m^{(0)}=\int\limits_{-2\pi/\sqrt{3}}^{2\pi/\sqrt{3}} \frac{\Lambda\left[\mathcal{Q}\left(\kappa,\alpha \right) \left(1+2\cos\left(\sqrt{3}\alpha/2\right)e^{-3i\kappa/2}\right)-\left(M_1\Omega_0^2-3 \right)\mathcal{P}\left(\kappa,\alpha \right) \right]e^{i m\left[3\kappa + \sqrt{3}\alpha \right]/2}}{\left[4\cos\left(3\kappa/2\right)\cos\left(\sqrt{3}\alpha/2\right)+2\cos\left(\sqrt{3}\alpha\right)-6+6M_1\Omega_0^2-M_1^2\Omega_0^4\right]}d\alpha,
\label{eq:hon_defect_P_IFT_eqn}\eeq 
\beq
\hspace{-1cm} Q_m^{(0)}=\int\limits_{-2\pi/\sqrt{3}}^{2\pi/\sqrt{3}} \frac{\Lambda\left[\mathcal{P}\left(\kappa,\alpha \right) \left(1+2\cos\left(\sqrt{3}\alpha/2\right)e^{3i\kappa/2}\right)-\left(M_1\Omega_0^2-3 \right)\mathcal{Q}\left(\kappa,\alpha \right) \right]e^{i m\left[3\kappa + \sqrt{3}\alpha \right]/2}}{\left[4\cos\left(3\kappa/2\right)\cos\left(\sqrt{3}\alpha/2\right)+2\cos\left(\sqrt{3}\alpha\right)-6+6M_1\Omega_0^2-M_1^2\Omega_0^4\right]}d\alpha,
\label{eq:hon_defect_Q_IFT_eqn}\eeq 
\beq
\hspace{-0.4cm}\Lambda=\left(M_1-M_0\right)\frac{\sqrt{3}}{4\pi^2}\Omega_0^2, \hspace{0.2cm}\mathcal{P}(\kappa,\alpha)=P_0^{(0)}+P_1^{(0)}e^{-i\left[3\kappa + \sqrt{3}\alpha \right]/2},\hspace{0.2cm}\mathcal{Q}(\kappa,\alpha)=Q_0^{(0)}+Q_1^{(0)}e^{-i\left[3\kappa + \sqrt{3}\alpha \right]/2}.
\nonumber\eeq Due to the amount of algebra involved, from heron in we opt to focus solely on the newly formed standing wave frequency at $\kappa=\pi/3$ of the highest branch (figure \ref{fig:armchair_2scale}(b)). The algebra is more substantial than the zigzag defect section because each of the integrals \eqref{eq:hon_defect_P_IFT_eqn}, \eqref{eq:hon_defect_Q_IFT_eqn} need to be resolved for both $m=0$ and $1$ and solved. 
\\

It is worth noting that an alternative formulation of the armchair lattice is available whereby we consider an elementary cell consisting of 4 masses. The lattice vector ${\bf e_1}$ remains the same but ${\bf e_2}$ is redefined as the orthogonal vector $\sqrt{3}\epsilon {\bf j}$. In our current system's notation an example of the elementary cell would be $\left[q_{n,m}, p_{n,m}, p_{n,m+1},q_{n-1,m+1} \right]$. It follows that our original coordinate system (figure \ref{fig:armchair_lattice}) and the orthogonal system have identical $\kappa.{\bf e_1}$ values. This corresponds to identical short-scale phase shifts between masses in the ${\bf e_1}$ direction, hence the $\Omega(\kappa)$ dispersion curves for both the perfect and defect lattices will be identical for both formulations. This in turn allows us to derive a dispersion curve $\Omega\left(\kappa_1,\kappa_2 \right)$ for the defect-free lattice, along an easily retrievable Brillouin zone. The resulting irreducible Brillouin zone associated to our orthogonal system is rectangular (inset in figure \ref{fig:armchair_2scale}(a)) and it can be seen that the defect curve ($M_0 \neq M_1$) in figure \ref{fig:armchair_2scale}(b) is spawned from the defect-free curve ($M_0=M_1$), which is also present along the path $AB$ in figure \ref{fig:armchair_2scale}(a). As was the case for the zigzag defect a standing wave which shifts as $M_0$ decreases is present along the second highest branch of the dispersion curve (figure \ref{fig:defect_defect_zigzag}(b)). The appearance and disappearance at $M_0=0.5$ of which is attributed to the increasing localisation of the oscillations as $M_0$ tends to zero.

\vspace{0.5cm}
\indent We return to our asymptotics where at leading order for $M_1=1, M_0=0.4$ we find that the double eigenvalue under consideration is $\Omega_0=3.3643509309$ with associated eigenvectors ${\bf U}\hspace{-0.14cm}=\hspace{-0.14cm}[1, -0.7017854774, 0, 0.7123883445]^T, {\bf V}\hspace{-0.14cm}=\hspace{-0.14cm}[0,-0.7123883445, 1, -0.7017854774]^T$, such that the leading order displacement is 
\beq
{\bf P}^{(0)} = f^{(1)}(\eta) {\bf U}+ f^{(2)}(\eta) {\bf V}, \hspace{0.3cm} {\bf P}^{(0)}=\left[P^{(0)}_0, Q^{(0)}_0,P^{(0)}_1, Q^{(0)}_1 \right]^T.
\label{eq:armchair_displacement} \eeq  Finally at $\mathcal{O}(\epsilon)$ after implementing the solvability condition $\left[{\bf P}^{(0)}\right]^{T}\left[A^{(1)} \left(\Omega_1 \right) \right]{\bf P}^{(0)}=0$ we find the following coupled equations
\beq
f^{(2)}_{\eta}-\tau\Omega_1^2 f^{(1)}=0, \hspace{0.2cm} f^{(1)}_{\eta}+\tau\Omega_1^2 f^{(2)}=0, \hspace{0.3cm}\tau=0.3811310829.
\eeq If we apply the Bloch periodicity conditions to the long-scale displacements $f^{(1)}_{\eta}, f^{(2)}_{\eta}$ we deduce the first-order correction which can be seen to describe the local behaviour about $\kappa=\pi/3$, figure \ref{fig:armchair_2scale}(b). 

\subsubsection{Defect within the line-defect}
\begin{figure}[hb!]
\begin{center}
    \includegraphics[height=5.5cm, width=12cm]{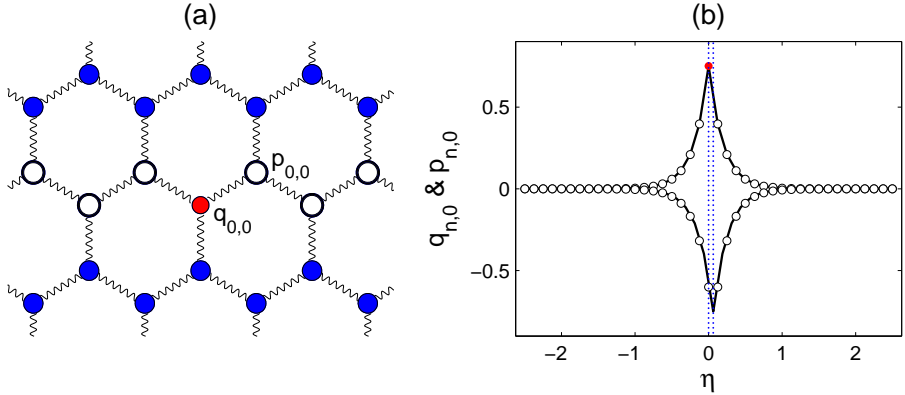}
\end{center}
\caption{\small {Panel (a) shows the defect within the embedded defect. Panel (b) shows a comparison between our asymptotic method (solid lines) and the mass positions found from the numerics (circles). The number of masses in either direction is taken as $N=41$ and the masses with the positive displacements are $q_{n,0}$ whilst the masses with the negative displacements are $p_{n,0}$. The asymptotic curve in the upper half-plane is $f^{(2)}(\eta)$ and the lower half-plane curve is $f^{(2)}(\eta-\epsilon/2)$. The location of the maximum displacement for $p_{n,0}$ and $q_{n,0}$ is indicated by the vertical dashed lines. For the mass values specified in the text we obtain the following frequency values: $\Omega=3.607958243$ (numerics) and $\Omega=3.607414703$ (asymptotics). $\eta=\epsilon n$ where $\epsilon=0.125$ and the decay rate $\beta=5.082845700$.   }}
\label{fig:defect_defect_zigzag}
\end{figure}

Here we shall consider the coupled system of equations associated to the honeycomb lattice where a single defect is introduced into the zigzag defect, figure \ref{fig:defect_defect_zigzag}(a). Previously in \cite{joseph13a}  the efficacy of our methodology was demonstrated for the uncoupled system with the square geometry, where we only dealt with a single displacement function. Hence we have omitted the hexagonal lattice case due to its similarities with the square geometry.

\vspace{0.5cm}

\indent We leave equation \eqref{eq:zigzag_difference_equations1} unchanged and alter equation \eqref{eq:zigzag_difference_equations2} whereby we multiply the defective mass term $M_0$ by the factor $\left(1+ \alpha \epsilon \delta_{n,0}\right)$. This term accounts for the mass change at the position associated to the displacement $q_{0,0}$ where the newly introduced defective mass has the value $M_{*}$ and where we redefine $\epsilon$ as $\epsilon=1-M_{*}/M_0 \ll 1$.  We still apply our two-scales procedure solely in the direction of zigzag defect, as the new mass alteration only takes affect at $\mathcal{O}\left(\epsilon^2 \right)$. For simplicity we shall only consider defective modes present within the stop-band of the Bloch diagram (figure \ref{fig:zigzag_disp_lattice}(a)) hence we consider locally in-phase behaviour between diagonal columns of masses, $p_m\left(\eta,0 \right)=p_m\left(\eta,\pm 1\right)$ (similar for $q_m\left(\eta,0 \right)$). It follows that in order to obtain a mode within the desired band we require that $M_1>M_0>M_*$ which in turn implies that $\alpha=-1$ . Previously for the line-defect we assumed that the long-scale modulation for both masses contained within the elementary cell of a honeycomb lattice were identical. However due to the asymmetric nature of our newly introduced defect, we now assume that  $p_m^{(0)}=f^{(1)}(\eta)P_m^{(0)}$, $q_m^{(0)}=f^{(2)}(\eta)Q_m^{(0)}$ where $f^{(1)}(\eta)=f^{(2)}(\eta-\epsilon/2)$. This motivates us to find the correct equation governing motion which is dependent only on the function $f^{(2)}(\eta)$. Note that our assumption is visually justified by observing the location of the defect within the masses, as shown in figure \ref{fig:defect_defect_zigzag}(a).
\\

The asymptotic procedure is identical at leading order to the $\kappa=0$ case, outlined in the previous section, whilst at $\mathcal{O}(\epsilon)$ we now obtain the relation $p_0^{(1)}=-q_0^{(1)}$. This change, at an order lower than that of the defect, is due to the Taylor's expansion of  $f^{(2)}(\eta-\epsilon/2)$ at leading order. Finally at $\mathcal{O}\left(\epsilon^2 \right)$, for the mass values $M_1=1, M_0=0.4, M_*=0.35$ we deduce the following ODE 
\beq
f^{(2)}_{\eta,\eta}+\tau_1 \Omega_2^2f^{(2)} -\tau_2 \delta(\eta) f^{(2)}=0, \hspace{0.25cm} \tau_1=(0.6097781588)^{-1},\hspace{0.25cm} \tau_2=10.16569140.
\eeq The above equation is expected to concede a solution of the form $f^{(2)}=\exp\left(-\beta|\eta| \right)$, where the decay rate $\beta$, along with $\Omega_2^2$, is to be found. Equations involving the two unknowns are found by examining the ODE for the case $\eta \neq 0$ and by employing the following continuous Fourier transforms,
\beq
 \tilde {f}(\gamma)=\int_{-\infty}^\infty f(\eta) e^{i\gamma \eta} d\eta,\qquad
f(\eta)=\frac{1}{2\pi}\int_{-\infty}^\infty \tilde{f}(\gamma) e^{-i\gamma \eta} d\gamma,
 \label{eq:continuousFT}
\eeq where eventually we find that $\beta^2=-\tau_1\Omega_2^2$ and $\beta=\tau_2/2$. 
 \\
 
 The accuracy of our asymptotics is verified against the numerics, which are formed by  altering the previous matrix equation \eqref{eq:honeycomb_matrix_equations2} into the following problem
\beq
\widehat{H}Q+Q\widehat{H}+EQE+E^TQE^T-\left(\mathcal{M}_1 Q \mathcal{M}_1 + \mathcal{M}_1 \mathcal{M}_2 Q \mathcal{M}_2 \right)=0, \hspace{0.2cm} \widehat{H}=G+T,
\label{eq:zigzagdefect_matrix}\eeq where $\mathcal{M}_1$ has diagonal entries containing $(M_1\Omega^2-3)$ except in the central position of the matrix where the entry is $(M_0\Omega^2-3)$ and $\mathcal{M}_2$ contains a single non-zero element, also in the central position which takes the value $\sqrt{\left(M_*-M_0 \right)\Omega^2}$. A comparison between the numerics and asymptotics for both $p_{n,m}$ and $q_{n,m}$ along the zigzag defect, is shown in figure \ref{fig:defect_defect_zigzag}(b).

\bibliographystyle{siam}
\bibliography{Different_Geometries}

\end{document}